\documentclass[letterpaper,twocolumn,10pt]{article}

\usepackage{usenix}
\usepackage[english]{babel}
\usepackage{lipsum}
\usepackage{caption}
\usepackage{subcaption}
\usepackage{graphicx}
\usepackage[hidelinks]{hyperref}
\usepackage{cite}
\usepackage[usenames,dvipsnames]{color}
\usepackage{listings}
\usepackage{booktabs}
\usepackage{array}
\usepackage{multirow,tabularx}
\usepackage{makecell}
\usepackage{paralist}
\usepackage{xspace}
\usepackage{amsmath}
\usepackage{tikz}
\usetikzlibrary{shadows,arrows,shapes}
\usepackage{comment}

\usepackage{lscape}
\usepackage{multicol}

\usepackage{moresize}

\usepackage{anyfontsize}

\usepackage{marginnote}

\usepackage{pifont}

\usepackage{xcolor, colortbl}

\usepackage{verbatimbox}

\usepackage{xspace}

\usepackage{microtype}

\usepackage[nolist]{acronym}

\pdfmapfile{+zi4.map}
\usepackage{zi4}

\newcolumntype{L}[1]{>{\raggedright\let\newline\\\arraybackslash\hspace{0pt}}m{#1}}
\newcolumntype{C}[1]{>{\centering\let\newline\\\arraybackslash\hspace{0pt}}m{#1}}
\newcolumntype{R}[1]{>{\raggedleft\let\newline\\\arraybackslash\hspace{0pt}}m{#1}}

\lstdefinelanguage{JavaScriptHTML}{
    keywords={typeof, new, true, false, catch, function, return, null, catch, switch, var, if, in, while, do, else, case, break, function},
    keywordstyle=\color{BrickRed}\bfseries,
    ndkeywords={class, export, boolean, throw, implements, import, this},
    ndkeywordstyle=\color{Gray}\bfseries,
    identifierstyle=\color{black},
    sensitive=false,
    comment=[l]{//},
    morecomment=[s]{/*}{*/},
    morecomment=[s]{<!--}{-->},
    commentstyle=\color{Green}\ttfamily,
    stringstyle=\color{red}\ttfamily,
    morestring=[b]',
    morestring=[b]"
}

\lstdefinestyle{JS}{language=JavaScriptHTML,
    extendedchars=true,
    basicstyle=\scriptsize\ttfamily,
    showstringspaces=false,
    showspaces=false,
    numberstyle=\tiny\color{black},
    numbers=left,
    numbersep=8pt,
    tabsize=2,
    breaklines=true,
    showtabs=false,
    frame=single,
    frameround=ffff,
    captionpos=b
}

\newcommand{\todo}[1]{\textcolor{Orange}{TODO: #1}}

\begin{acronym}

\acro{3DES}{Triple-DES}

\acro{AES}{Advanced Encryption Standard}
\acro{ALU}{Arithmetic Logic Unit}
\acro{AOT}{Ahead-of-Time}
\acro{API}{Application Programming Interface}
\acro{ARX}{Addition Rotation XOR}
\acro{ASIC}{Application Specific Integrated Circuit}
\acro{ASIP}{Application Specific Instruction-Set Processor}
\acro{AS}{Active Serial}

\acro{BNF}{Backus-Naur Form}
\acro{BRAM}{Block-Ram}

\acro{CBC}{Cipher Block Chaining}
\acro{CFB}{Cipher Feedback Mode}
\acro{CFG}{Control Flow Graph}
\acro{CISC}{Complex Instruction Set Computer}
\acro{CLB}{Configurable Logic Block}
\acro{CMP}{Chemical Mechanical Polishing}
\acro{COFF}{Common Object File Format}
\acro{COTS}{Commercial Off-The-Shelf}
\acro{CPA}{Correlation Power Analysis}
\acro{CPU}{Central Processing Unit}
\acro{CRC}{Cyclic Redundancy Check}
\acro{CTR}{Counter}

\acro{DC}{Direct Current}
\acro{DCFG}{Dynamic Control Flow Graph}
\acro{DES}{Data Encryption Standard}
\acro{DFA}{Differential Frequency Analysis}
\acro{DFT}{Discrete Fourier Transform}
\acro{DLL}{Dynamic Link Library}
\acro{DMA}{Direct Memory Access}
\acro{DNF}{Disjunctive Normal Form}
\acro{DPA}{Differential Power Analysis}
\acro{DSO}{Digital Storage Oscilloscope}
\acro{DSP}{Digital Signal Processing}
\acro{DUT}{Device Under Test}

\acro{ECB}{Electronic Code Book}
\acro{ECC}{Elliptic Curve Cryptography}
\acro{ECDH}{Elliptic Curve Diffie-Hellman}
\acro{EEPROM}{Electrically Erasable Programmable Read-only Memory}
\acro{EMA}{Electromagnetic Emanation}
\acro{EM}{electro-magnetic}

\acro{FFT}{Fast Fourier Transformation}
\acro{FF}{Flip Flop}
\acro{FI}{Fault Injection}
\acro{FIR}{Finite Impulse Response}
\acro{FPGA}{Field Programmable Gate Array}
\acro{FSM}{Finite State Machine}

\acro{GUI}{Graphical User Interface}

\acro{HDL}{Hardware Description Language}
\acro{HD}{Hamming Distance}
\acro{HF}{High Frequency}
\acro{HSM}{Hardware Security Module}
\acro{HW}{Hamming Weight}

\acro{IC}{Integrated Circuit}
\acro{IDU}{Instruction Decode Unit}
\acro{IO}[I/O]{Input/Output}
\acro{IOB}{Input Output Block}
\acro{IOT}[IoT]{Internet of Things}
\acro{IP}{Intellectual Property}
\acro{IR}{Instruction Register}
\acro{ISA}{Instruction Set Architecture}
\acro{IV}{Initialization Vector}

\acro{JIT}{Just-in-Time}
\acro{JS}{JavaScript}
\acro{JTAG}{Joint Test Action Group}

\acro{KAT}{Known Answer Test}

\acro{LFSR}{Linear Feedback Shift Register}
\acro{LSB}{Least Significant Bit}
\acro{LUT}{Look-up table}

\acro{MAC}{Message Authentication Code}
\acro{MILS}{Multiple Independent Levels of Security}
\acro{MIPS}{Microprocessor without Interlocked Pipeline Stages}
\acro{MMIO}{Memory Mapped \acl{IO}}
\acro{MSB}{Most Significant Bit}

\acro{NASA}{National Aeronautics and Space Administration}
\acro{NSA}{National Security Agency}
\acro{NVM}{Non-Volatile Memory}

\acro{OFB}{Output Feedback Mode}
\acro{OISC}{One Instruction Set Computer}
\acro{ORAM}{Oblivious Random Access Memory}

\acro{PAR}{Place-and-Route}
\acro{PCB}{Printed Circuit Board}
\acro{PC}{Program Counter}
\acro{PLA}{Programmable Logic Array}
\acro{PS}{Passive Serial}
\acro{PUF}{Physical Unclonable Function}

\acro{RAM}{Random Access Memory}
\acro{SEM}{Scanning Electron Microscope}
\acro{RISC}{Reduced Instruction Set Computer}
\acro{RNG}{Random Number Generator}
\acro{ROM}{Read-Only Memory}
\acro{ROI}{Region Of Interest}
\acro{ROP}{Return-oriented Programming}
\acro{RTL}{Register Transfer Level}

\acro{SCA}{Side-Channel Analysis}
\acro{SHA}{Secure Hash Algorithm}
\acro{SNR}{Signal-to-Noise Ratio}
\acro{SPA}{Simple Power Analysis}
\acro{SPI}{Serial Peripheral Interface Bus}
\acro{SRAM}{Static Random Access Memory}

\acro{UART}{Universal Asynchronous Receiver Transmitter}
\acro{UHF}{Ultra-High Frequency}
\acro{USB}{Universal Serial Bus}

\acro{VHDL}{Very High Speed Integrated Circuit Hardware Description Language}

\acro{WISC}{Writeable Instruction Set Computer}

\acro{XTS}{XEX-based Tweaked-codebook with ciphertext Stealing}

\end{acronym}

\begin{document}

\title{Reverse Engineering x86 Processor Microcode}

\author{
    {\rm Philipp Koppe, Benjamin Kollenda, Marc Fyrbiak, Christian Kison,} \\ {\rm Robert Gawlik, Christof Paar, and Thorsten Holz}\\[1ex]
Ruhr-Universit\"at Bochum
}

\maketitle


\begin{abstract}

Microcode is an abstraction layer on top of the physical components of a CPU and
present in most general-purpose CPUs today. In addition to facilitate complex
and vast instruction sets, it also provides an update mechanism that allows CPUs
to be patched in-place without requiring any special hardware. While it is
well-known that CPUs are regularly updated with this mechanism, very little is
known about its inner workings given that microcode and the update mechanism are
proprietary and have not been throughly analyzed yet.

In this paper, we reverse engineer the microcode semantics and inner workings of
its update mechanism of conventional COTS CPUs on the example of AMD's K8 and
K10 microarchitectures. Furthermore, we demonstrate how to develop custom
microcode updates. We describe the microcode semantics and additionally present
a set of microprograms that demonstrate the possibilities offered by this
technology. To this end, our microprograms range from {CPU}-assisted
instrumentation to microcoded Trojans that can even be reached from within a web
browser and enable remote code execution and cryptographic implementation
attacks.
\end{abstract}


\section{Introduction}

Similar to complex software systems, bugs exist in virtually any commercial
\ac{CPU} and can imply severe consequences on system security, e.g., privilege
escalation~\cite{esorics:2008:duflot,asplos:2015:hicks} or leakage of
cryptographic keys~\cite{crypto:2008:biham}. Errata sheets from embedded to
general-purpose processors list incorrect behavior with accompanying workarounds
to safeguard program execution~\cite{intel:errata,amd:errata}. Such workarounds
contain instructions for developers on how these bugs can be bypassed or
mitigated, e.g., by means of recompilation~\cite{dsn:2008:meixner} or binary
re-translation~\cite{woar:2006:reis}. However, these interim solutions are not
suited for complex design errors which require hardware
modifications~\cite{iccd:2006:narayanasamy}. Dedicated hardware units to counter
bugs are imperfect~\cite{asplos:2015:hicks,micro:2007:sarangi} and involve
non-negligible hardware costs~\cite{micro:1999:austin}. The infamous
\textit{Pentium fdiv} bug~\cite{eetimes:1997:wolfe} illustrated a clear economic
need for field updates after deployment in order to turn off defective parts and
patch erroneous behavior. Note that the implementation of a modern processor
involves millions of lines of HDL code~\cite{sun:opensparc:t2} and verification
of functional correctness for such processors is still an unsolved
problem~\cite{intel:errata,amd:errata}.

Since the 1970s, x86 processor manufacturers have used microcode to decode
complex instructions into series of simplified microinstructions for reasons of
efficiency and diagnostics~\cite{tc:1980:rauscher}. From a high-level
perspective, microcode is an interpreter between the user-visible \ac{CISC}
\ac{ISA} and internal hardware based on \ac{RISC}
paradigms~\cite{book:computer_architecture:stallings}. Although microcode was
initially implemented in read-only memory, manufacturers introduced an update
mechanism by means of a patch \ac{RAM}.

Once erroneous \ac{CPU} behavior is discovered, manufacturers publish a
microcode update, which is loaded through the BIOS/UEFI or operating system
during the boot process. Due to the volatility of the patch \ac{RAM}, microcode
updates are not persistent and have to be reloaded after each processor reset.
On the basis of microcode updates, processor manufacturers obtain flexibility
and reduce costs of correcting erroneous behavior. Note that both Intel and AMD
deploy a microcode update mechanism since Pentium Pro (P6) in
1995~\cite{tr:2014:chen, intel:pprospec} and K7 in 1999~\cite{tr:2014:chen,
amd:revisionguide}, respectively. Unfortunately, \ac{CPU} vendors keep
information about microcode secret. Publicly available documentation and patents
merely state vague claims about how real-world microcode \textit{might} actually
look like, but provide little other insight.

\par{\bf Goals.} In this paper, we focus on microcode in x86 \acp{CPU} and our
goal is to answer the following research questions:

\begin{enumerate} \item What is microcode and what is its role in x86 \acp{CPU}?
\item How does the microcode update mechanism work? \item How can the
proprietary microcode encoding be reverse engineered in a structured,
semi-automatic way? \item How do real-world systems profit from microcode and
how can malicious microcode be leveraged for attacks? \end{enumerate}

In order to answer question~(1), we emphasize that information regarding
microcode is scattered among many sources (often only in patents). Hence, an
important part of our work is dedicated to summarize this prerequisite knowledge
forming the foundation to answer the more in-depth research questions.
Furthermore, we tackle shortcomings of prior attempted security analyses of x86
microcode, which were not able to reverse engineer microcode~\cite{tr:2014:chen,
link:2004:opteron_exposed}. We develop a novel technique to reverse engineer the
encoding and thus answer question~(2). After we obtain a detailed understanding
of the x86 microcode for several \ac{CPU} architectures, we can address
question~(3). As a result, we obtain an understanding of the inner workings of
\ac{CPU} updates and can even generate our own updates. In particular, we focus
on potential applications of microprograms for both defensive and offensive
purposes to answer question~(4). We demonstrate how a microprogram can be
utilized to instrument a binary executable on the \ac{CPU} layer and we also
introduce different kinds of backdoors that are enabled via microcode updates.

Our analysis focuses on the AMD K8/K10 microarchitecture since these \acp{CPU}
do not use cryptographic signatures to verify the integrity and authenticity of
microcode updates. Note that Intel started to cryptographically sign microcode
updates in 1995~\cite{tr:2014:chen} and AMD started to deploy strong
cryptographic protection in 2011~\cite{tr:2014:chen}. We assume that the
underlying microcode update mechanism is similar, but cannot analyze the
microcode updates since we cannot decrypt them.

\par{\bf Contributions.} In summary, our main contributions in this paper are as
follows: \begin{itemize} \item {\bf In-depth Analysis of Microcode.} We provide
an in-depth overview of the opaque role of microcode in modern \acp{CPU}. In
particular, we present the fundamental principles of microcode updates as
deployed by vendors to patch \ac{CPU} defects and errors.

\item {\bf Novel RE Technique.} We introduce the first semi-automatic reverse
engineering technique to disclose microcode encoding of general-purpose
\acp{CPU}. Furthermore, we describe the design and implementation of our
framework that allows us to perform this reverse engineering.

\item {\bf Comprehensive Evaluation.} We demonstrate the efficacy of our
technique on several \ac{COTS} AMD x86 \ac{CPU} architectures. We provide the
microcode encoding format and report novel insights into AMD x86 \ac{CPU}
internals. Additionally, we present our hardware reverse engineering findings
based on delayering actual \acp{CPU}.

\item {\bf Proof-of-Concept Microprograms.} We are the first to present
fully-fledged microprograms for x86 \acp{CPU}. Our carefully chosen
microprograms highlight benefits as well as severe consequences of unveiled
microcode to real-world systems.

\end{itemize}


\section{Related Work}

Before presenting the results of our analysis process, we briefly review existing literature on microprogramming and related topics.

\par{\bf Microprogramming.}
Since Wilkes' seminal work in 1951~\cite{mucic:1951:wilkes}, numerous works in academia as well as industry adopted and advanced microprogrammed \ac{CPU} designs. Diverse branches of research related to microprogramming include higher-level microcode languages, microcode compilers and tools, and microcode verification~\cite{book:1976:agrawala,tc:1980:rauscher,cav:2005:arons}. Other major research areas focus on optimization of microcode, i.e., minimizing execution time and memory space~\cite{ieee_cs:1975:jones}. In addition, several applications of microprogramming were developed~\cite{micro:1974:habib} such as diagnostics~\cite{micro:1988:melvin}.

Since microcode of today's x86 \acp{CPU} has not been publicly documented yet, several works attempted a high-level security analysis for \acp{CPU} from both Intel and AMD~\cite{tr:2014:chen, link:2004:opteron_exposed}. Even though these works reported the workings of the microcode update mechanism, the purpose of fields within the microcode update header, and the presence of other metadata, none of the works was able to reverse engineer the essential microcode encoding. Hence, they were not able to build microcode updates on their own.

We want to note that Arrigo Triulzi presented at TROOPERS'15 and '16 that he had been able to patch the microcode of an AMD K8 microarchitecture~\cite{Arrigo:2015:Troopers, Arrigo:2016:Troopers}. However, he did neither publish the details of his reverse engineering nor the microcode encoding.

\par{\bf Imperfect \ac{CPU} Design.}
Although microcode updates can be leveraged to rectify some erroneous behavior, it is not a panacea. Microcode updates are able to degrade performance due to additional condition checks and they cannot be applied in all cases. An infamous example is AMD's K7, where the microcode update mechanism itself was defective~\cite{tr:2014:chen, amd:revisionguide}.
In order to tackle these shortcomings, diverse techniques have been proposed including dynamic instruction stream editing~\cite{isca:2003:corliss}, field-programmable hardware~\cite{micro:2007:sarangi}, and hardware checks~\cite{micro:1999:austin,asplos:2015:hicks}.

\par{\bf Trusted Hardware.}
The security of applications and operating systems builds on top of the security of the underlying hardware. Typically software is not designed to be executed on untrusted or potentially malicious hardware~\cite{link:2007:raadt,crypto:2008:biham,esorics:2008:duflot}. Once hardware behaves erroneously (regardless of whether deliberately or not), software security mechanisms can be invalidated. Numerous secure processors have been proposed over the years~\cite{isca:2005:suh,ccs:2013:maas,usenix:2016:costan}. Commercially available examples include technologies such as Intel SGX~\cite{iacr:2016:86} and AMD Pacifica~\cite{amd:pacifica}.

However, the periodicity of security-critical faults~\cite{intel:errata,amd:errata} and undocumented debug features~\cite{esorics:2008:duflot} in closed-source \ac{CPU} architectures challenges their trustworthiness~\cite{iacr:2016:86, link:2015:rutowska}.


\section{Microcode}

As noted earlier, microcode can be seen as an abstraction layer on top of the physical components of a \ac{CPU}. In this section, we provide a general overview of the mechanisms behind microcode and also cover details about the microcode structure and update mechanism.

\subsection{Overview}

The \ac{ISA} provides a consistent interface to software and defines instructions, registers, memory access, I/O, and interrupt handling. This paper focuses on the x86 \ac{ISA}, and to avoid confusion, we refer to x86 instructions as \emph{macroinstructions}. The microarchitecture describes how the manufacturer leveraged processor design techniques to implement the \ac{ISA}, i.e., cache size, number of pipeline stages, and placement of cells on the die. From a high-level perspective, the internal components of a processor can be subdivided into data path and control unit. The data path is a collection of functional units such as registers, data buses, and \ac{ALU}. The control unit contains the \ac{PC}, the \ac{IR} and the \ac{IDU}. The control unit operates diverse functional units in order to drive program execution. More precisely, the control unit translates each macroinstruction to a sequence of actions, i.e., retrieve data from a register, perform a certain \ac{ALU} operation, and then write back the result. The \textit{control signal} is the collection of electrical impulses the control unit sends to the different functional unit in one clock cycle. The functional units produce \textit{status signals} indicating their current state, i.e., whether the last \ac{ALU} operation equals zero, and report this feedback to the control unit. Based on the status signals, the control unit may alter program execution, i.e., a conditional jump is taken if the zero flag is set.

The \ac{IDU} plays a central role within the control unit and generates control signals based on the contents of the instruction register. We distinguish between two \ac{IDU} implementation concepts: (1) hardwired and (2) microcoded.

\par{\bf Hardwired Decode Unit.}
A hardwired decode unit is implemented through sequential logic, typically a \ac{FSM}, to generate the instruction-specific sequence of actions. Hence, it provides high efficiency in terms of speed. However, for complex \acp{ISA} the lack of hierarchy in an \ac{FSM} and  state explosion pose challenging problems during the design and test phases~\cite{book:practical-introduction-hardware-software-codesign}. Hardwired decode units inhibit flexible changes in the late design process, i.e., correcting bugs that occurred during test and verification, because the previous phases have to be repeated. Furthermore, post-manufacturing changes (to correct bugs) require modification of the hardware, which is not (economically) viable for deployed \acp{CPU}~\cite{eetimes:1997:wolfe}. Hence, hardwired decode units are suited for simple \acp{ISA} such as \ac{RISC} processors like SPARC and MIPS.

\par{\bf Microcoded Decode Unit.}
In contrast to the hardwired approach, the microcoded \ac{IDU} does not generate the control signals on-the-fly, but rather replays precomputed \textit{control words}. We refer to one control word as \emph{microinstruction}. A microinstruction contains all control information required to operate all involved functional units for one clock cycle. We refer to a plurality of microinstructions as \emph{microcode}. Microinstructions are fetched from the microcode storage, often implemented as on-chip \ac{ROM}. The opcode bytes of the currently decoded macroinstruction are leveraged to generate an initial address, which serves as the entry point into microcode storage. Each microinstruction is followed by a \emph{sequence word}, which contains the address to the next microinstruction. The sequence word may also indicate that the decoding process of the current macroinstruction is complete. It should be noted that one macroinstruction often issues more than one microinstruction. The microcode sequencer operates the whole decoding process, successively selecting microinstructions until the \emph{decode complete} indicator comes up. The microcode sequencer also handles conditional microcode branches supported by some microarchitectures. Precomputing and storing control words introduces flexibility: Changes, patches, and adding new instructions can be moved to the late stages of the design process. The design process is simplified because changes in decode logic only require adaption of the microcode \ac{ROM} content. On the downside, decoding latency increases due to \ac{ROM} fetch and multistage decode logic. A microcoded \ac{IDU} is the prevalent choice for commercial \ac{CISC} processors.

\subsection{Microcode Structure}

Two common principles exist to pack control signals into microinstructions. This choice greatly influences the whole microarchitecture and has a huge impact on the size of microcode programs.

\par{\bf Horizontal Encoding.}
The horizontal encoding designates one bit position in the microinstruction for each control signal of all functional units. For the sake of simple logic and speed, no further encoding or compression is applied. This results in broad control words, even for small processors. The historical IBM System/360 M50 processor with horizontally-encoded microcode used 85-bit control words~\cite{link:2012:micropr_history}. The nature of horizontal microcode allows the programmer to explicitly address several functional units at the same time to launch parallel computations, thus using the units efficiently. One disadvantage is the rather large microcode \ac{ROM} due to the long control words.

\par{\bf Vertical Encoding.}
Vertically encoded microcode may look like a common \ac{RISC} instruction set. The microinstruction usually contains an opcode field that selects the operation to be performed and additional operand fields. The operand fields may vary in number and size depending on the opcode and specific flag fields. Bit positions can be reused efficiently, thus the microinstructions are more compact. The lack of explicit parallelism simplifies the implementation of microcode programs, but may impact performance. One encoded operation may activate multiple control signals to potentially several functional units. Hence, another level of decoding is required. The microcode instruction set and encoding should be chosen carefully to keep the second-level decoding overhead minimal.

\subsection{Microcode Updates}
\label{subsec:micro_update}

One particular benefit of microcoded microarchitectures is the capability to install changes and bug fixes in the late design process. This advantage can be extended even further: With the introduction of microcode updates, one can alter processor behavior even after production. Manufacturers leverage microcode patches for debugging purposes and fixing processor errata. The well-known \textit{fdiv} bug~\cite{eetimes:1997:wolfe}, which affected Intel Pentium processors in 1994, raised awareness that similarly to software, complex hardware is error-prone, too. This arguably motivated manufacturers to drive forward the development of microcode update mechanisms. Typically, a microcode patch is uploaded to the \ac{CPU} by the motherboard firmware (e.g., BIOS or UEFI) or the operating system during the early boot process. Microcode updates are stored in low-latency, volatile, on-chip RAM. Consequently, microcode patches are not persistent. Usually, the microcode patch \ac{RAM} is fairly limited in size compared to microcode \ac{ROM}. A microcode patch contains a number of microinstructions, sequence words, and triggers. Triggers represent conditions upon which the control is transferred from microcode \ac{ROM} to patch \ac{RAM}. In a typical use case, the microcode patch intercepts the \ac{ROM} entry point of a macroinstruction. During instruction decode, the microcode sequencer checks the triggers and redirects control to the patch RAM if needed. A typical microcode program residing in patch \ac{RAM} then may, for example, sanitize input data in the operands and transfer control back to the microcode \ac{ROM}.


\section{Reverse Engineering Microcode}

In this section, we provide an overview of the AMD K8 and K10 microarchitecture families and describe our reverse engineering approach. Furthermore, we present our analysis setup and framework that includes prototype implementations of our concepts and supported our reverse engineering effort in a semi-automated way.

Our analysis primarily covers AMD K8 and K10 processors because---to the best of our knowledge---they are the only commercially available, modern x86 microarchitectures lacking strong cryptographic protection of microcode patches.

\subsection{AMD K8 and K10}

AMD released new versions of its K8 and K10 processors from 2003 to 2008 and 2008 to 2013, respectively. Note that the actual production dates may vary and in 2013 only two low-end CPU models with K10 architecture were released. K9 is the K8's dual-core successor, hence the difference is marginal from our point of view. Family 11h and 12h are adapted K10 microarchitectures for mobile platforms and APUs.

All of theses microarchitectures include a microcoded IDU. The x86 instruction set is subdivided into \textit{direct path} and \textit{vector path} macroinstructions. The former mainly represent the frequently used, performance critical macroinstructions (e.g., arithmetic and logical operations) that are decoded by hardware into up to three microinstructions. The latter are uncommon or complex, and require decoding by the microcode sequencer and microcode \ac{ROM}. Vector path macroinstructions may produce many microinstructions. During execution of the microcode sequencer, hardware decoding is paused. The microcode is structured in \textit{triads} of three 64-bit  microinstructions and one 32-bit sequence word~\cite{tr:2014:chen}. An example microinstruction set is described in AMD's patent RISC86~\cite{patent:2002:risc86} from 2002. The sequence word may contain the address of the next triad or indicate that decoding is complete. The microcode \ac{ROM} is addressed in steps whose length is a triad. An example address space ranging from \texttt{0x0} to \texttt{0xbff} thus contains 3,072 triads. The microcode is responsible for the decoding of vector path macroinstructions and handling of exceptions, such as page faults and divide-by-zero errors.

\subsection{Update Mechanism}
\label{ucode:section:reverse-engineering-microcode:update-mechanism}

The K7, released in 1999, was AMD's first microarchitecture supporting microcode updates. The update mechanism did not change throughout to the 12h family. AMD kept the update feature secret until it was exposed along with three K8 microcode patches in 2004. The patches and the update mechanism were reverse engineered from BIOS updates~\cite{link:2004:opteron_exposed}. The microcode updates are stored in a proprietary file format, although pieces of information have been reverse engineered~\cite{tr:2014:chen, link:2004:opteron_exposed}. With the K10 microarchitecture, AMD started to publicly release microcode updates, which benefits the Linux open-source microcode update driver. Our view of the file format is depicted in Table~\ref{fig:ucode_update_file_format} including the header with checksum and number of triads, match register fields, and triads. It should be noted that triads in microcode updates are obfuscated with an algorithm we do  not specify further due to ethical considerations.

\begin{table}[!htb]
    \centering
    \resizebox{0.995\linewidth}{!}{
\begin{tabular}{|l|cccccccccccccccccccccccccccccccccccccccccccccccccccccccccccccccc|}
\hline
\cellcolor{black!15}B$\downarrow$ Bit$\rightarrow$& \multicolumn{32}{|c|}{\cellcolor{black!15}0\hspace{85pt}31} & \multicolumn{32}{|c|}{\cellcolor{black!15}32\hspace{85pt}63}\\
\Xhline{2\arrayrulewidth}
\cellcolor{black!15} 0 & \multicolumn{32}{|c}{date} & \multicolumn{32}{|c|}{patch ID}\\
\hline
\cellcolor{black!15} 8 & \multicolumn{16}{|c}{patch block} & \multicolumn{8}{|c}{len} & \multicolumn{8}{|c}{init} & \multicolumn{32}{|c|}{checksum}\\
\hline
\cellcolor{black!15} 16 & \multicolumn{32}{|c}{northbridge ID} & \multicolumn{32}{|c|}{southbridge ID}\\
\hline
\cellcolor{black!15} 24 & \multicolumn{32}{|c}{CPUID} & \multicolumn{32}{|c|}{magic value}\\
\hline
\cellcolor{black!15} 32 & \multicolumn{32}{|c}{match register 0} & \multicolumn{32}{|c|}{match register 1}\\
\hline
\cellcolor{black!15} 40 & \multicolumn{32}{|c}{match register 2} & \multicolumn{32}{|c|}{match register 3}\\
\hline
\cellcolor{black!15} 48 & \multicolumn{32}{|c}{match register 4} & \multicolumn{32}{|c|}{match register 5}\\
\hline
\cellcolor{black!15} 54 & \multicolumn{32}{|c}{match register 6} & \multicolumn{32}{|c|}{match register 7}\\
\hline
\cellcolor{black!15} 64 & \multicolumn{64}{|c|}{triad 0, microinstruction 0}\\
\hline
\cellcolor{black!15} 72 & \multicolumn{64}{|c|}{triad 0, microinstruction 1}\\
\hline
\cellcolor{black!15} 80 & \multicolumn{64}{|c|}{triad 0, microinstruction 2}\\
\hline
\cellcolor{black!15} 88 & \multicolumn{32}{|c}{triad 0, sequence word} & \multicolumn{32}{|c|}{triad 1 ...}\\
\hline
\end{tabular}
}
    \caption{Microcode update file format.}
    \label{fig:ucode_update_file_format}
\end{table}

\par{\bf Microcode Update Procedure.}
The microcode update binary is uploaded to the \ac{CPU} in the following way: First, the patch must be placed in accessible virtual address space. Then the 64-bit virtual address must be written to Model-Specific Register (MSR) \texttt{0xc0010020}. Depending on the update size and microarchitecture, the \texttt{wrmsr} instruction initiating the update may take around 5,000 cycles to complete. Rejection of a patch causes a general protection fault. Internally, the update mechanism verifies the checksum, copies the triads to microcode patch RAM, and stores the match register fields in the actual match registers. Patch RAM is mapped into the address space of the microcode ROM, whereby the patch triads directly follow the read-only triads.

\par
\begin{figure}[!htb]
	\centering
	\resizebox{0.995\linewidth}{!}{
		\includegraphics{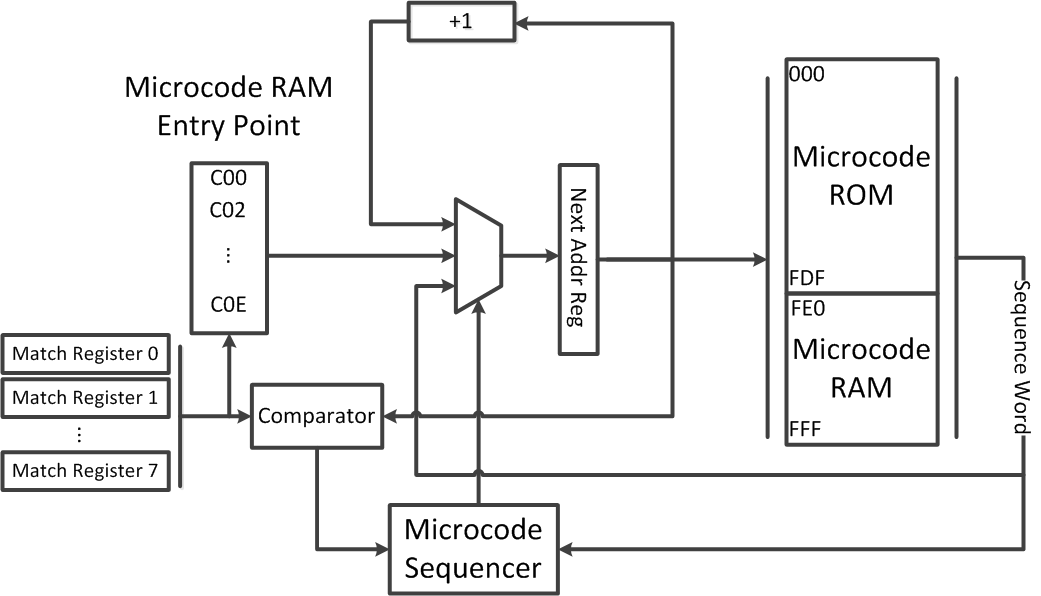}
	}
	\caption{Overview ofthe  AMD microcode update mechanism.}
	\label{ucode:figure:microcode:update}
\end{figure}

{\bf Match Registers.}
The match registers are an integral part of the update mechanism. They hold a microcode ROM address, intercept the triad stored at that location, and redirect control to the triad in patch RAM at the offset \emph{match register index $\cdot$ 2}. The shared address space enables microcode in the patch RAM to jump back to microcode ROM, e.g., to reuse existing triads. Due to the complexity of the microcode update procedure we assume it is implemented in microcode itself. We summarize our understanding of the microcode update mechanism in~Figure~\ref{ucode:figure:microcode:update}. AMD's patent~\cite{patent:2002:patch_device} from 2002 describes an example microcode patch device and provides an idea of how the internals work.

\subsection{Reverse Engineering Methods}
\label{subsect:re_methods}

Based on our insights into microcode and its update mechanism, we now detail our novel method used to reverse engineer the microcode encoding. More precisely, we employ a (1)~low-noise environment as a foundation for the novel (2)~microcode \ac{ROM} heat map generation, and (3)~the microcode encoding reverse engineering. Furthermore, we present (4)~microcode hooking which ultimately enables actual modification of \ac{CPU} behavior.

We would like to emphasize that our methods were developed when we did not have access to microcode \ac{ROM}, see Section~\ref{sec:hw_re}.

\par{\bf Low-Noise Environment.}
Since we did not have access to \ac{CPU} internals, we had to be able to apply our crafted microcode updates and carefully analyze the  modified \ac{CPU}'s behavior (e.g., register values and memory locations). To pinpoint exactly where the changes caused effects (down to a single macroinstruction), we had to eliminate any \textit{noise} from parallel or operating system code executions out of our control. For example, common operating systems implement task switching or fully symmetric multiprocessing, which is undesirable in our setting. This code execution is capable of triggering abnormal behavior (because of our microcode update) and then most likely causes a system crash.
Hence, we require a \textit{low-noise} environment where we have full control of all code to realize accurate observation of the \ac{CPU} state and behavior.

\par{\bf Microcode \ac{ROM} Heat Maps.}
As described in Section~\ref{ucode:section:reverse-engineering-microcode:update-mechanism}, match registers hold microcode \ac{ROM} addresses. Since we did not know which microcode \ac{ROM} addresses belong to which macroinstructions, we were not able to change the behavior for a specific microcoded macroinstruction. Hence, we devised \textit{microcode \ac{ROM} heat maps}, a method to discover the corresponding memory location for microcoded macroinstructions.

The underlying idea is to generate distinct behavior between the original and the patched macroinstruction execution. More precisely, the patch contains a microcode instruction that always crashes on execution. Thereby, we generate a \textit{heat map} for each macroinstruction in an automated way: we store whether the microcode \ac{ROM} address causes a system crash or not. The comparison between original and patched execution reveals which microcode \ac{ROM} addresses correspond to the macroinstruction. We further automatically processed all heat maps to exclude common parts among all macroinstructions.

\par{\bf Microcode Encoding Reverse Engineering.}
Based on our automatically generated heat maps, we were able to tamper with a specific microcoded macroinstruction. However, we could not meaningfully alter an instruction because of its proprietary encoding. Hence, we developed a novel technique to reverse engineer proprietary microcode encoding in a semi-automatic way.

Since we did not have a large microcode update base on which we could perform fine-grained tests, we merely had a black box model of the \ac{CPU}. However, since microinstructions control \ac{ALU} and register file accesses, we formed various general assumptions about the instruction fields, which can be systematically tested using semi-automatic tests (e.g., opcode, immediate value, source and destination register fields).

In order to reverse engineer the encoding, we applied a two-tiered approach.
First, we identified fields by means of bits that cause similar behavior, i.e., change of used registers, opcode, and immediate value. Second, we exhaustively brute-forced each field to identify all addressable values. Since corresponding fields are small ($< 10$ bits), we combined the results together and gradually formed a model of the encoding. Note that through detailed exception reporting and paging, we were able to gather detailed information on why a specific microinstruction caused a crash. Earlier in the reverse engineering process, we set the three microinstructions in a triad to the same value to avoid side effects from other unknown microinstructions. Once we had a better understanding of the encoding, we padded the triad with no-operation microinstructions. Later in the reverse engineering process, we designed tests that reuse microinstructions from existing microcode updates. For that method to be successful, a good understanding of the operand fields was required as most of these microinstructions operate on internal registers. We had to rewrite the register fields to be able to directly observe the effect of the microinstruction. Furthermore, we designed automated tests that identified set bits in unknown fields of existing microinstructions and permuted the affected bit locations in order to provoke observable differences in behavior that can be analyzed.

\par{\bf Microcode Hooks.}
After reverse engineering the microcode encoding, we can arbitrarily change \ac{CPU} behavior for any microcoded macroinstruction and intercept control for any microcode \ac{ROM} address. Note that we intercepted a macroinstruction at the entry point microcode \ac{ROM} address. In order to realize a fully-fledged \textit{microcode hook} mechanism, we have to correctly pass back control after interception through our microcode update. This is indispensable in case macroinstructions are extended with functionality, such as a conditional operand check, while preserving original functionality.

We employed two basic concepts to resume macroinstruction computation after interception: (1) pass control back to \ac{ROM}, and (2) implement the macroinstruction computation. Note that we implemented both resume strategies, see Section~\ref{ucode:section:microprograms}.

\subsection{Framework}

One fundamental requirement for our framework was automated testing. Combined with the fact that microcode updates potentially reset or halt the entire machine, it became apparent that another controller computer was needed. In the following, we describe both our hardware setup and our framework implementation.

\par{\bf Hardware Setup.}
From a high-level point of view, the hardware setup consists of multiple nodes and several development machines. Each node represents one minimal computer with an AMD \ac{CPU} that runs our low-noise environment and is connected to a Raspberry Pi via serial bus. To enable monitoring and control, the mainboard's power and reset switch as well as the power supply's +3.3V are connected to GPIO ports. The Raspberry Pis run Linux and can be remotely controlled from the Internet. The development machines are used to design test cases and extend the microcode API. Furthermore, test cases can be launched from the development machines. This process automatically transfers the test case and the latest API version to the desired nodes, which then autonomously execute the test case and store the results. Our test setup consists of three nodes with K8 Sempron 3100+ (2004), K10 Athlon II X2 260 (2010), and K10 Athlon II X2 280 (2013) processors.

\par{\bf Low-Noise Environment.}
To fulfil our unusual requirements regarding the execution environment (e.g., full control over interrupts and all code being executed), we implemented a simple operating system from scratch. It supports interrupt and exception handling, virtual memory, paging, serial connection, microcode updates, and execution of streamed machine code. The streamed machine code serves the purpose of bringing the \ac{CPU} to the desired initial state, executing arbitrary macroinstructions, and observing the final state of the \ac{CPU}. We leveraged this feature primarily to execute vector path instructions intercepted by a microcode patch. This way, we can infer the effects of triads, single microinstructions, and the sequence word. Note that only the final state can be observed in case no exception occurs.

We implemented interrupt and exception handling in order to observe the intermediate state of the \ac{CPU} and the exception code such as general protection faults. The error state includes the faulting program counter and stack pointer as well as the x86 general-purpose registers. We refined the preciseness of the error reporting by implementing virtual memory and paging support. All exceptions related to memory accesses raise page faults with additional information such as the faulting address and action. This information, paired with the information about the faulting program counter, allows us to distinguish between invalid read, write, and execution situations. We also used the exception code and observed the intermediate state to infer the effects of microcode. A custom message protocol exposes the following operating system features via serial connection: (1) stream x86 machine code, (2) send and apply microcode update, and (3) report back the final or intermediate \ac{CPU} state. Some of the test processors support \textit{x86\_64 long mode}, which lets the \ac{CPU} access 64-bit instructions and registers. However, our operating system runs in \textit{32-bit protected mode}.

\par{\bf Microcode API.}
Our controller software is implemented in Python and runs on the Raspberry Pis. It processes test cases in an automated fashion and makes heavy use of the microcode API. Test cases contain an initial \ac{CPU} state, arbitrary x86 instructions, the final \ac{CPU} state, and an exception information filter plus a logger as well as a high-level microcode patch description. The microcode patch is generated with the high-level microcode patch information that includes header fields, match register values, and microcode in the form of bit vectors, \ac{RTL} machine language, or a mix. Test cases incorporating automation must specify at least one property that will be altered systematically. For example, a test case that aims to iteratively intercept all triads in microcode ROM may increment the match register value in each pass. Another test case that attempts to infer conditional behavior of microcode may alter streamed x86 machine code in order to induce different x86 eflags register values and at the same time permute the bit vector of an unknown field within a microinstruction. The microcode API exposes all required underlying features such as serial connection handling, serial message protocol, AMD computer power state monitoring and control, x86 assembler, parsing and generation of microcode updates, obfuscation and deobfuscation of microcode updates, microcode assembler and disassembler as well as required data structures. The framework runs through 190 test iterations per minute and node in case there are no faults. One fault adds a delay of 12 seconds due to the reboot.


\section{Microcode Specification}

In this section we present the results of our reverse engineering effort such as heat maps, a detailed description of the microcode instruction set, and intercepting x86 instructions. Furthermore, we present our microcode \ac{RTL}.

\par{\textsc{Disclaimer.}}
It should be noted that our results originate from reverse engineering include and indirectly measured behavior, assumptions about the microarchitecture, and interpretation of the visible \ac{CPU} state, which is small in comparison to the whole unobservable \ac{CPU} state. Thus, we cannot guarantee that our findings are intended behavior of AMD's microcode engine.

\subsection{Heat Maps}
\label{ucode:results:heatmaps}
A heat map of a specific macroinstruction contains a mapping of all microcode ROM addresses to a boolean value that indicates whether the specified triad is executed during the decode sequence of that macroinstruction. During the test cycle, our operating system executes vector instructions such as \texttt{call} and \texttt{ret}. We name a heat map that only covers vector instructions from the operating system \textit{reference heat map}. In order to obtain a clean heat map for a vector instruction, the reference heat map must be subtracted from the instruction's raw heat map. For the interested reader we present a truncated, combined K10 heat map in Table~\ref{fig:heatmap} in Appendix~\ref{ucode:appendix:microcode_specification}. The heat maps represent a fundamental milestone of our reverse engineering effort. They indicate microcode ROM locations to intercept macroinstructions and help infer logic from triads. We designed test cases for all vector path instructions, which then generated clean heat maps in a fully automated way.

\subsection{Microcode Instruction Set}
\label{ucode:results:ucode_instruction_set}
The microinstruction set presented in AMD's patent RISC86~\cite{patent:2002:risc86} gave us a general understanding and valuable hints. However, we found that almost all details such as microinstruction length, operand fields, operations, and encoding differ. Furthermore, we could not confirm that single microinstructions can be addressed, which would result in the preceding microinstructions of the triad being ignored. Instead, we found that only entire triads are addressable.
In the following, we reuse terminology from the patent where appropriate. Unless stated otherwise, all information given afterwards was  obtained through reverse engineering.

We found four operation classes, namely \textsf{RegOp}, \textsf{LdOp}, \textsf{StOp}, and \textsf{SpecOp}, that are used for arithmetic and logic operations, memory reads, memory writes, and special operations such as write program counter, respectively. The structure of the four operation classes is shown in Table~\ref{fig:opclass_encodings}. The different operation classes can be distinguished by the \textsf{op class} field at bit locations 37 to 39. \textsf{RegOp} and \textsf{SpecOp} share the same \textsf{op class} field encoding but have disjunct encodings for the operation type field. The unlabeled fields indicate unused or unknown bit locations. \textsf{RegOp} supports operation types such as arithmetic, comparators, and logic operations. The \texttt{mul} and \texttt{imul} operation types must be the first microinstruction within a triad in order to work.
\textsf{SpecOp} enables to write the x86 program counter and to  conditionally branch to microcode. If the conditional branch is taken, the microcode sequencer continues decoding at the given address. In case the conditional branch is not taken, the sequence word determines further execution. The condition to be evaluated is encoded in the 4 high bits of the 5-bit \textsf{cc} field. Bit 0 of the \textsf{cc} field inverts the condition if set. The available condition encodings match the ones given in patent RISC86~\cite{patent:2002:risc86}, p. 37. The \textit{write-program-counter} \textsf{SpecOp} must be placed third within a triad in order to work. We found that \textsf{LdOp} and \textsf{StOp} have their own operation types. Our collection of operation types is incomplete, because it was impossible to observe the internal state of the CPU. We show encoding details for the operation types we found in the Appendix in Table~\ref{fig:op_types}. The fields \textsf{reg1}, \textsf{reg2} and \textsf{reg3} encode the microcode registers. In addition to the general-purpose registers, microcode can access a number of internal registers. Their content is only stored until the microinstruction has been decoded.
The special \texttt{pcd} register is read-only and contains the address of the next macroinstruction to decode. This is valuable information to implement relative x86 jumps in microcode. The microarchitecture also contains a microcode substitution engine, which automatically replaces operand fields in the microinstruction with operands from the macroinstruction. The first two x86 operands can be accessed in microcode with the register encodings \texttt{regmd} and \texttt{regd}. We refer to Table~\ref{fig:reg_encodings} in the Appendix for encoding details of the microcode registers. We did not find the substitution mechanism for immediate values encoded in the macroinstruction. To solve this issue, we read the x86 instruction bytes from main memory and extract the immediate. The \textsf{sw} field swaps source and destination registers. The \textsf{3o} field enables the three operand mode and allows \textsf{RegOp} microinstructions of the form \texttt{reg2:=~reg1~op~reg3/imm}. The \textsf{flags} field decides whether the resulting flags of the current \textsf{RegOp} microinstruction should be committed to the x86 flags register. The \textsf{rmod} field switches between \texttt{reg3} and a 16-bit immediate value.
The sequence word, see Table~\ref{fig:sw_encoding}, contains an \textsf{action} field at bit locations 14 to 16 that may indicate a branch to the triad at the given \textsf{address}, a branch to the following triad, or stop decoding of the current macroinstruction. Our disassembler has a coverage of approximately 40\% of the instructions contained in existing microcode patches. However, we ignored bits in unknown fields of recognised microinstructions whose meaning we could not determine. We designed automated test cases that, e.g., permute the bits of an unknown microinstruction field to provoke observable differences in the final CPU state. Our result filter discarded outputs that match the expected CPU state. We then manually inspected the remaining interesting CPU states and inferred the meaning of the new encoding.

\begin{table*}[!htb]
    \centering
    \resizebox{0.995\linewidth}{!}{
\begin{tabular}{|l|cccccccccccccccccccccccccccccccccccccccccccccccccccccccccccccccc|}
\hline
\cellcolor{black!15} Index & \cellcolor{black!15} 63 & \multicolumn{9}{|c}{\cellcolor{black!15}62\hspace{17pt}54} & \multicolumn{1}{|c|}{\cellcolor{black!15}53} & \multicolumn{1}{c|}{\cellcolor{black!15}52} & \multicolumn{6}{c|}{\cellcolor{black!15}51\hspace{5pt}46} & \multicolumn{6}{c|}{\cellcolor{black!15}45\hspace{30pt}40} & \multicolumn{3}{c|}{\cellcolor{black!15}39\hspace{5pt}37} & \multicolumn{7}{c|}{\cellcolor{black!15}36\hspace{10pt}30} & \multicolumn{6}{c|}{\cellcolor{black!15}29\hspace{5pt}24} & \cellcolor{black!15}23 & \multicolumn{7}{|c|}{\cellcolor{black!15}22\hspace{5pt}16} & \multicolumn{16}{c|}{\cellcolor{black!15}15\hspace{45pt}0}\\
\Xhline{2\arrayrulewidth}
\cellcolor{black!15} \text{RegOp} &  -    & \multicolumn{9}{|c}{type} &                            \multicolumn{1}{|c}{sw} & \multicolumn{1}{|c}{3o} & \multicolumn{6}{|c}{reg1} & \multicolumn{3}{|c|}{-} & \multicolumn{2}{c}{flags} & \multicolumn{1}{|c|}{-} & \multicolumn{3}{c|}{000} & \multicolumn{4}{c|}{-} & \multicolumn{3}{c|}{size} & \multicolumn{6}{c|}{reg2} & \multicolumn{1}{c|}{rmod} & \multicolumn{7}{c|}{-} & \multicolumn{16}{c|}{imm16/reg3}\\
\hline
\cellcolor{black!15} \text{LdOp} &   -    & \multicolumn{9}{|c}{type} &                            \multicolumn{1}{|c}{sw} & \multicolumn{1}{|c}{3o} & \multicolumn{6}{|c}{reg1} & \multicolumn{6}{|c|}{-} &                                                     \multicolumn{3}{c|}{001} & \multicolumn{7}{c|}{-} &                               \multicolumn{6}{c|}{reg2} & \multicolumn{1}{c|}{rmod} & \multicolumn{7}{c|}{-} & \multicolumn{16}{c|}{imm16/reg3}\\
\hline
\cellcolor{black!15} \text{StOp} &   -    & \multicolumn{9}{|c}{type} &                            \multicolumn{1}{|c}{sw} & \multicolumn{1}{|c}{3o} & \multicolumn{6}{|c}{reg1} & \multicolumn{6}{|c|}{-} &                                                     \multicolumn{3}{c|}{010} & \multicolumn{4}{c|}{-} & \multicolumn{3}{c|}{size} & \multicolumn{6}{c|}{reg2} & \multicolumn{1}{c|}{rmod} & \multicolumn{7}{c|}{-} & \multicolumn{16}{c|}{imm16/reg3}\\
\hline
\cellcolor{black!15} \text{SpecOp} &  -    & \multicolumn{4}{|c}{type} & \multicolumn{5}{|c}{cc} & \multicolumn{1}{|c}{sw} & \multicolumn{1}{|c}{3o} & \multicolumn{6}{|c}{reg1} & \multicolumn{6}{|c|}{-} &                                                     \multicolumn{3}{c|}{000} & \multicolumn{4}{c|}{-} & \multicolumn{3}{c|}{size} & \multicolumn{6}{c|}{reg2} &                             \multicolumn{8}{c|}{-} & \multicolumn{16}{c|}{imm16/addr12}\\
\hline
\end{tabular}
}
    \caption{The four operation classes and their microinstruction encoding.}
    \label{fig:opclass_encodings}
\end{table*}

\begin{table}[!htb]
    \centering
    \resizebox{0.995\linewidth}{!}{
\begin{tabular}{|l|cccccccccccccccccccccccccccccccc|}
\hline
\cellcolor{black!15} Index & \multicolumn{15}{|c}{\cellcolor{black!15}31\hspace{35pt}17} & \multicolumn{3}{|c}{\cellcolor{black!15}16\hspace{5pt}14} & \multicolumn{2}{|c|}{\cellcolor{black!15}13\hspace{5pt}12} & \multicolumn{12}{c|}{\cellcolor{black!15}11\hspace{30pt}0}\\
\Xhline{2\arrayrulewidth}
\cellcolor{black!15} \text{next\_triad} & \multicolumn{15}{|c}{-} & \multicolumn{3}{|c}{000} & \multicolumn{14}{|c|}{-}\\
\hline
\cellcolor{black!15} \text{branch} & \multicolumn{15}{|c}{-} & \multicolumn{3}{|c}{010} & \multicolumn{2}{|c}{-} & \multicolumn{12}{|c|}{address}\\
\hline
\cellcolor{black!15} \text{complete} & \multicolumn{15}{|c}{-} & \multicolumn{3}{|c}{110} & \multicolumn{14}{|c|}{-}\\
\hline
\end{tabular}
}
    \caption{Sequence word encoding.}
    \label{fig:sw_encoding}
\end{table}

\subsection{Intercepting x86 Instructions}

Currently, we can only intercept vector instructions by writing related triad addresses from the heat maps into the match registers. We are uncertain whether a mechanism for hooking direct path instructions exists. It is relatively simple to replace the logic of a vector path instruction; however, it appeared challenging to \textit{add} logic, because the original semantics must be preserved.
To solve this issue, we leverage the two microcode hook concepts from Section~\ref{subsect:re_methods}. In the following we describe in detail the practical application of both concepts. (1) After executing the added logic, we jump back to microcode ROM. (2) After execution of the added logic we implement the semantics of the macroinstruction in microcode ourselves and indicate \textit{sequence complete} in the last triad. This way, we successfully hooked \textit{shrd} and \textit{imul} vector path instructions.

We also successfully intercepted the \textit{div} instruction using the first method. One fundamental limitation of hooking with match registers is that one cannot jump back to the intercepted triad, because the match register would redirect control again, essentially creating an endless loop. We are not aware of a feature to temporarily ignore a match register. Thus we need to intercept a negligible triad and, after execution of our logic, jump back to the subsequent triad, essentially skipping one triad. We inferred the observable part of the logic of div heat map triads. We proceeded by iteratively branching directly to the triads with a known \ac{CPU} state with a match register hook set to the following triad. With this method we found one triad we can skip without visibly changing the result. Specifically, we can intercept triad \texttt{0x7e5} per match register, induce the desired behavior, and finally jump back to address \texttt{0x7e6} via sequence word. It should be noted that the hook is in the middle of the calculation. Thus the source and destination general-purpose registers as well as some internal microcode registers hold intermediate results, which need to be preserved if the correctness of the final result matters.

\subsection{Microcode RTL}

We developed a microcode register transfer language based on the syntax of Intel x86 assembly language, because for the implementation of microprograms it is impractical to manually assemble bit vectors. In the following, we show a template for a typical microinstruction in our microcode \ac{RTL}:
\begin{center}
\texttt{insn op1, op2[, op3]}
\end{center}
The \textsf{insn} field defines the operation. It is followed by one to three operands of which the first one is always the destination and only the last one may be an immediate. In two-operand mode, the first operand is the destination and the source. There are dedicated load and store instructions. Memory addressing currently supports only one register, i.e., \texttt{ld~eax,~[ebx]}. The size of arithmetic operations is implicitly specified by the destination operand's size. Memory reads always fetch a whole native system word,  and the size of memory writes is specified by the source operand's size. The conditional microcode branch encodes the condition in the first operand and the branch target in the second operand, i.e. \texttt{jcc~nZF,~0xfe5}.
The assembler automatically resolves constraints such as \texttt{mul} must be placed first in a triad and \textit{write-program-counter} must be placed last. Strictly speaking the sequence words are not instructions, thus we cover them by directives such as \texttt{.sw\_complete} and \texttt{.sw\_branch~0x7e6}. The \textit{branch to next triad} sequence words are added implicitly.


\section{Hardware Analysis}
\label{sec:hw_re}

In addition to the black box microcode reverse engineering presented in the previous section, we analyzed the \ac{CPU}'s hardware in a parallel approach. The goal of hardware analysis was to read and analyze the non-volatile microcode \ac{ROM} to support reverse engineering of the microcode encoding. Furthermore, this allows us to analyze the actual implementation of microcoded macroinstructions.

Our chosen \ac{DUT} is a Sempron 3100+ (SDA3100AIP3AX) with a 130nm technology size, since it features the largest size of the target \ac{CPU} family (which facilitates our analysis). Note that the larger technology size allows for additional tolerance margins in both the delayering and the imaging of the individual structures.
Similar to any common microcontroller or \ac{CPU}, the \ac{DUT} is built using a CMOS process with multiple layers. In contrast to traditional microcontrollers, general-purpose x86 \acp{CPU} feature a much larger die size and are stacked up to 12 layers, which increases hardware reverse engineering effort.

We expected the targeted non-volatile microcode \ac{ROM} to be stored in a cell array architecture. Other memory types to implement microcode \ac{ROM}, such as flash, \ac{EEPROM}, and \ac{RAM}, are either too slow, unnecessarily large, or volatile.\\
Note that the general die structure is almost identical to the die shot provided in \cite{vries2003_chip_architect}, which helped our initial analysis identify our \ac{ROI}, the microcode \ac{ROM}.

\subsection{Delayering}
After removing the heat sink with a drill, we fully decapsulated the die with fuming nitric acid~\cite{jetcs:2016:quadir}.
In order to visualize the \ac{ROM} array, we \emph{delayered} (e.g., removed individual stacked layers) from the top of the die. The main challenge during delayering is to uniformly skim planar surfaces parallel to the individual layers. Typically, the delayering process alternates between removing a layer and imaging the layer beneath it~\cite{jetcs:2016:quadir}. Focusing on our \ac{ROI}, we were able to neglect other areas of the chip resulting in a more planar surface in important region(s). Note that hardware reverse engineering of the whole \ac{CPU} microarchitecture would require a more controlled delayering process and several months to acquire and process the whole layout. The interested reader is referred to our die shot in Figure~\ref{fig:cpu_overview} in the Appendix.

In order to remove layers, we used a combined approach of \ac{CMP} and plasma etching. During inspection of the seventh layer, we encountered the expected \ac{ROM} array structure. We acquired images of individual layers using a \ac{SEM} since optical microscopy reaches diffraction limits at this structure size. Compared to colored and more transparent images from optical microscopy, \ac{SEM} images only provide a gray-scale channel, but with higher magnification. In \ac{SEM} images, different materials can be identified due to brightness yield.

We encountered multiple regular NOR \ac{ROM} arrays using contact layer (vias) for programming. In NOR \ac{ROM} with active layer programming, the logic state is encoded by the presence or absence of a transistor~\cite{Skorobogatov05semi-invasiveattacks}. In our case an advanced \textit{bitline-folding} architecture~\cite{Jacob:2007:MSC:1543376} encodes the logic state by either placing a via on the right or the left bitline. Note another property of this \ac{ROM} type is that only a single via may be set at any time; setting both will result in a short circuit.

Overall, we identified three \ac{ROM} blocks consisting of 8 subarrays. Each of the 3 \ac{ROM} blocks has the capability to store 30 kB. Note that our results match the visible blocks in~\cite{vries2003_chip_architect}.
It is important to note that the vias' positions are hardwired and cannot be changed after shipping. The only possible way to patch bugs in the \ac{ROM} is to employ the microcode update procedure described in Section~\ref{subsec:micro_update}.

\subsection{Microcode Extraction}

\begin{figure}[!h]
        \centering
        \resizebox{1\columnwidth}{!}{
                \includegraphics{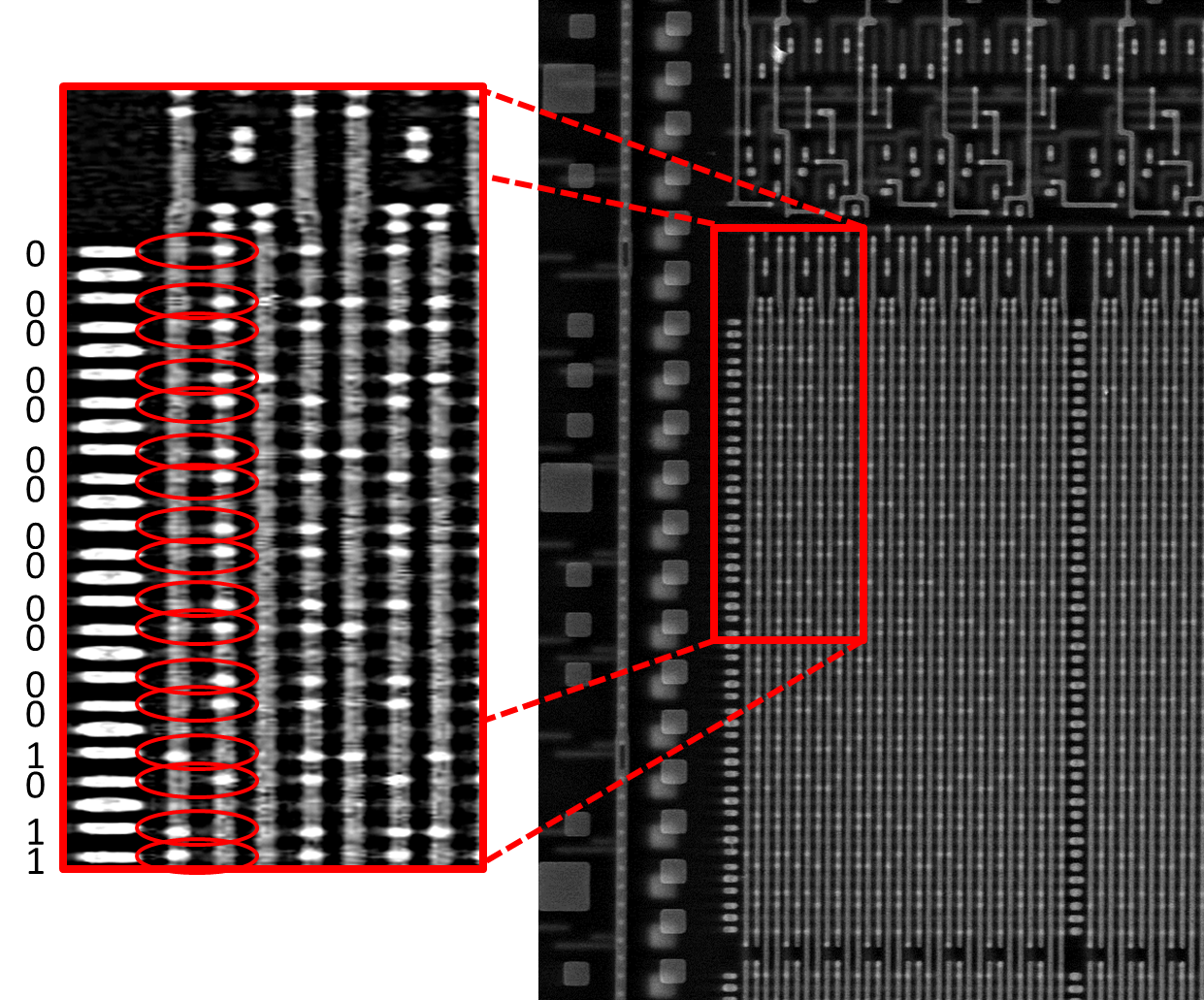}
          }
        \caption{Partially interpreted bits in one ROM subarray.}
        \label{ucode:figure:rom}
\end{figure}

In Figure~\ref{ucode:figure:rom}, we highlighted how bits are programmed by this memory type. Bright spots represent a via going down from a metal line, which is either connected to GND or VCC.
We chose to represent the individual cells as set to logical '1' if the left via was set and '0' if the right one was set. This convention does not necessarily correspond to the correct runtime interpretation. However, permutations are commonly applied to the \ac{ROM} memory, hence a misinterpretation can be corrected in a later analysis step.

In order to analyze the microcode \ac{ROM} bits for any permutations, we processed the acquired \ac{SEM} images with \textit{rompar}~\cite{rompar}. Using its image processing capabilities, we transformed the optical via positions into bit values.

\par{\bf Microcode ROM Bit Analysis.}
In order to group the bit values into microinstructions, we carefully analyzed the \ac{ROM} structure and we made two crucial observations:
(1) Each alternating column of bits is inverted due to mirroring of existing cells, which saves space on the die.
(2) Since the memory type employs a transposed bitline architecture~\cite{Jacob:2007:MSC:1543376}, the bit inversion has to be adjusted to each segment.

With both observations in mind, we were able to derive microinstructions from the images. Note that we also had to interleave the subarrays respectively to acquire 64 bits (size of a microinstruction) per memory row. Hence, the \ac{ROM} allows us to find more complex microinstructions and experimentally reverse engineer their meaning.


\section{Microprograms}
\label{ucode:section:microprograms}

In this section, we demonstrate the effectiveness of our reverse engineering effort by presenting microprograms that successfully augment existing x86 instructions and add foreign logic. With this paper, we also publish microcode patches~\cite{microcode:amd_microprograms} that are compiled from scratch and run on unmodified AMD CPUs, namely K8 Sempron 3100+ and K10 Athlon II X2 260/280. We found that the microcode ROM content varies between different processors, but the macroinstruction entry points into the microcode ROM are constant. Thus we assume our microcode patches are compatible with a wider range of K8/K10-based CPUs. We discuss additional applications of microcode in Section~\ref{sec:discussion}.

\subsection{Instrumentation}
\label{ucode:section:microprograms:instrumentation}

Instrumentation monitors the execution of a program and may produce metadata or instruction traces. It is used by program analysis, system defenses, antivirus software, and performance optimization during software development. It has been proven challenging to implement performant instrumentation for \ac{COTS} binaries. Several mechanisms exist such as function hooking, binary rewriting, virtual machine introspection, and in-place emulation. However, they come with drawbacks such as coarse granularity, uncertain coverage, and high performance overhead. An instrumentation framework with \ac{CPU} support based on microcode may evade many of the disadvantages. It should be noted that microcode also has limitations such as only 8 match registers. Currently we can only intercept vector path x86 instructions and the hooks are machine-wide, i.e., not limited to one user-space process. For demonstration purposes we implemented a simple instrumentation that counts the occurrences of the \texttt{div} instruction during execution. See Listing~\ref{ucode:listing:instrumentation_simple} for a high-level representation of the instrumentation logic; we refer the interested reader to Listing~\ref{ucode:listing:instrumentation} in Appendix~\ref{ucode:appendix:microprograms} for a detailed \ac{RTL} implementation.

\begin{lstlisting}[
	label=ucode:listing:instrumentation_simple,
	aboveskip=1.5em,
	caption={High-level description of the instrumentation logic implemented in microcode that counts the \texttt{div} instructions during execution.}
]
if (esi == magic) {
    temp = dword [edi]
    temp += 1
    dword [edi] = temp
}
\end{lstlisting}

\subsection{Remote Microcode Attacks}
\label{sec:subsec:ucode_programs:remote}
Executing microcode Trojans is not limited to a local attacker. An injected
microcode hook may lie dormant within a vector path macroinstruction, such as a
\texttt{div reg32}, and it is triggered as soon as a specific \emph{trigger}
condition is met
within an attacker-controlled web page. This is possible due to \ac{JIT}
and \ac{AOT} compilers embedded in modern web browsers.
They allow to emit specific machine code instructions only utilizing \ac{JS}.
Consider a microcode Trojan for the \texttt{div}
instruction. We provide a high-level description of the Trojan logic in Listing~\ref{ucode:listing:div_trojan:asm}.
\begin{lstlisting}[
	label=ucode:listing:div_trojan:asm,
	aboveskip=1.5em,
	caption={High-level description of the microcode Trojan implemented in microcode that increments the \texttt{eip} to execute x86 instructions in a disaligned fashion.}
]
if (eax == A && ebx == B)
    eip = eip + 1
\end{lstlisting}
If a \texttt{div ebx} instruction is executed while \texttt{eax} contains the
value \texttt{A} (dividend) and \texttt{ebx} contains the value \texttt{B}
(divisor), then the instruction pointer \texttt{eip} is increased, and execution
continues in a misaligned way after the first byte of the instruction following the
\texttt{div ebx} instruction. If the trigger condition is not met, the division
is executed as expected.
Hence, legitimate machine instructions as shown in Listing~\ref{ucode:listing:asm_js_native_excerpt:asm} may be misused to hide and execute arbitrary code.
\begin{minipage}{\linewidth}
\begin{lstlisting}[
	label=ucode:listing:asm_js_native_excerpt:asm,
	aboveskip=1.5em,
	caption={x86 machine code to trigger the \texttt{div} Trojan in Listing~\ref{ucode:listing:div_trojan:asm}.}
]
B8 0A000000    mov eax, 0xA
BB 0B000000    mov ebx, 0xB
F7F3           div ebx
05 909090CC    add eax, 0xCC909090
\end{lstlisting}
\end{minipage}

Due to the microcode Trojan within \texttt{div ebx}, which is triggered when the condition \texttt{eax == A \&\& ebx == B} is met, the instruction following the division is
executed starting at its second byte (Listing~\ref{ucode:listing:asm_js_native_payload:asm}).

\begin{lstlisting}[
	label=ucode:listing:asm_js_native_payload:asm,
	aboveskip=1.5em,
    caption={x86 hidden payload executed due to the triggered microcode Trojan.}
]
B8 0A000000    mov eax, 0xA
BB 0B000000    mov ebx, 0xB
F7F3           div ebx
05             /* SKIPPED */
90             nop
90             nop
90             nop
CC             int3
\end{lstlisting}
As shown in Listing~\ref{ucode:listing:asm_js_native_payload:asm}, the hidden \texttt{nop} and \texttt{int3} instructions
within the constant value of the \texttt{add} instruction are
executed instead of the legitimate \texttt{add} itself. Note that many
\texttt{add} instructions can be used to hide an arbitrary payload (i.e.,
\texttt{execve()}) instead of \texttt{nop} and \texttt{int3}. 

We were able to emit appropriate machine code instructions using the \emph{ASM.JS} subset of the \ac{JS} language in Mozilla Firefox 50. ASM.JS compiles a web page's \ac{JS} code before it is actually transformed into native machine code. We hide
our payload within four-byte \ac{JS} constants of legitimate
instructions similar to previous \ac{JIT} Spraying
attacks~\cite{blazakis2010interpreter,sintsov2010jit}. Since we also
control the dividend and divisor of the division, we eventually trigger the microcode
Trojan in the \texttt{div} instruction,
which in turn starts to execute our payload. Thus, we achieved to remotely activate the microcode hook and use it to execute remotely controlled machine code. We refer the interested reader to the ASM.JS code in Listing~\ref{appendix:listing:asm.js_trigger} in Appendix~\ref{appendix:subsec:asm.js_trigger}.
While usually \emph{constant blinding} is used in \ac{JIT} compilers to prevent the
injection of valid machine code into \ac{JS} constants, recent research has shown
that browsers such as Microsoft Edge or Google Chrome fail to blind constants in
certain cases~\cite{Dachshund2017}. Hence, we assume that remotely triggering a microcode Trojan
and executing hidden code within other browsers
(i.e., Edge or Chrome) is possible, too. 

\subsection{Cryptographic Microcode Trojans}
\label{ucode:section:ucode_programs:crypto}

In order to demonstrate further severe consequences of microcode Trojans, we detail how such Trojans facilitate implementation attacks on cryptographic algorithms. More precisely, we present how microcode Trojans enable both (1) a bug attack (representative for \ac{FI}~\cite{eurocrypt:1997:boneh}) and (2) a timing attack for \ac{SCA}~\cite{crypto:1996:kocher}.

\subsubsection{Preliminaries and Goal}
\ac{ECC} has become the prevalent public-key cryptographic primitive in real-world systems. In particular, numerous cryptographic libraries, e.g., OpenSSL and libsodium, employ \textsf{Curve25519}~\cite{pkc:2006:bernstein}. Note that the critical scalar multiplication is generally implemented through a Montgomery ladder whose execution is expected to be constant time, see RFC7748~\cite{rfc7748}.

\par{\bf Bug Attack.}
Bug attacks~\cite{crypto:2008:biham,ctrsa:2012:brumley} are associated with \ac{FI}; however, they are conceptionally distinct. While \ac{FI} mainly considers faults injected by an adversary, bug attacks rely on inherent computation bugs\cite{ches:2016:ghandali} and do neither suppose environmental tampering nor physical presence.

\par{\bf Timing Attack.}
Timing attacks~\cite{crypto:1996:kocher} against cryptographic implementations are based on careful analysis of their execution time~\cite{usenix:2003:brumley,cans:2016:kaufmann}. Nowadays most libraries employ constant-time implementations as an effective countermeasure.

Our goal for each attack is to enable  disclosure of the private key from \acs{ECDH} key exchange. In order to realize microcode Trojans which facilitate such attacks, we have to arm a microcoded x86 instruction (used in scalar multiplication) with (1) an input-dependent trigger and (2) a payload inducing a conditional fault or additional time, see Listing~\ref{ucode:listing:crypto:trojan}.

\begin{lstlisting}[
	label=ucode:listing:crypto:trojan,
	aboveskip=1.5em,
	caption={High-level microcode Trojan description within an x86 instruction to trigger a conditional bug using the first operand (\texttt{regmd}) of the x86 instruction and the immediate constants \texttt{A} and \texttt{C}.}
]
if (regmd == A)
    regmd = regmd + C
\end{lstlisting}

\subsubsection{Implementation}

For both attacks, we use the constant-time ECC reference implementation from libsodium~\cite{libsodium:reference_implementation} compiled for 32-bit architectures.
Since \textsf{Curve25519} employs reduced-degree reduced-coefficient polynomials for arithmetic and the implementation uses 64-bit data types, the following C code 
is compiled to assembly in Listing~\ref{ucode:listing:crypto:asm}:
\begin{verbbox}
	carry = (h + (i64) (1L << 25)) >> 26;
\end{verbbox}
\begin{center}
	\theverbbox
\end{center}

\begin{minipage}{\linewidth}
\begin{lstlisting}[
	label=ucode:listing:crypto:asm,
	aboveskip=1.5em,
	caption={x86 machine code implementing 64-bit right shift using the \texttt{shrd} instruction.}
]
	mov    eax, dword [esp+0xd0]
	add    eax, 0x2000000
	mov    ebx, dword [esp+0xd4]
	adc    ebx, 0x0
	shrd   eax, ebx, 0x1a
\end{lstlisting}
\end{minipage}

This line of code processes internal (key-dependent) data as well as adversary-controlled (public-key dependent) data. We can remotely trigger the condition in the microcoded \texttt{shrd} instruction to apply both the bug attack and the timing attack. Note that in case of a timing-attack, we conditionally execute several \texttt{nop} instructions to induce a data-dependent timing difference.

For a detailed \ac{RTL} implementation of the bug attack, we refer the interested reader to Listing~\ref{ucode:listing:crypto:ucode_trojan} in Appendix~\ref{ucode:appendix:microprograms}. We emphasize that the necessary primitives for bug attacks and timing side channel attacks can be created via microcode Trojans. This way, even state-of-the-art cryptographic implementations can be undermined.

\section{Discussion}
\label{sec:discussion}

\subsection{Security Implications}

We demonstrated that malware can be implemented in microcode. Furthermore, malicious microcode updates can be applied to unmodified K8 and K10-based AMD \acp{CPU}. This poses a certain security risk.
However, in a realistic attack scenario, an adversary must overcome other security measures. A remote attacker has to bypass application and operating system isolation in order to apply a microcode update. An attacker with system privileges might as well leverage less complex mechanisms with better persistence and stealth properties than microcode malware.
An attacker with physical access may be able to embed a malicious microcode update into the BIOS or UEFI, i.e., in an \textit{evil maid} scenario~\cite{link:evil_maid}. However, she has to overcome potential security measures such as TPM or signing
 of the UEFI firmware. Physical access also enables alternative attack vectors such as cloning the entire disk, or in case of full disk encryption, tamper with the MBR or bootloader.
Other adversary models to provide malicious microcode (either through updates or directly in microcode \ac{ROM}) become more realistic, i.e., intelligence agencies or untrusted foundries. From a hardware Trojan's perspective~\cite{ieee_dt:2010:tehranipoor}, microcode Trojans provide post-manufacturing versatility, which is indispensable for the heterogeneity in operating systems and applications running on general-purpose \acp{CPU}.

Even though AMD emphasizes that their chips are secure~\cite{http://www.fudzilla.com/home/item/32120-amd-denies-existence-of-nsa-backdoor}, the microcode update scheme of K8 and K10 shows once more that \textit{security by obscurity} is not reliable and proper encryption, authentication, and integrity have to deployed.\\
It should be noted that attacks leveraging microcode will be highly hardware-specific. Current AMD processors employ strong cryptographic algorithms to protect the microcode update mechanism~\cite{tr:2014:chen}. Microcode and its effects on system security for current \acp{CPU} are unknown with no verifiable trust anchor. Both experts and users are unable to examine microcode updates for (un)intentional bugs.

\subsection{Constructive Microcode Applications}

We see great potential for constructive applications of microcode in \ac{COTS} \acp{CPU}. We already discussed that microcode combines many advantages for binary instrumentation, see Section~\ref{ucode:section:microprograms:instrumentation}. This could aid program tracing, bug finding, tainting, and other applications of dynamic program analysis. Furthermore, microcode could boost the performance of existing system defenses. Microcode updates could also enable domain-specific instruction sets, e.g., special instructions that boost program performance or trustworthy security measures (similar to Intel SGX~\cite{iacr:2016:86}).

Hence, the view on microcode and its detailed embedding in the overall \ac{CPU} architecture are a relevant topic for future research.

\subsection{Generality}

In addition to x86 \ac{CISC} \acp{CPU} from Intel, AMD, and VIA, microcode is also used in \acp{CPU} based on \ac{RISC} methodologies.
For example, reverse engineering of an ARM1 processor~\cite{link:reverse-engineering-arm1} disclosed the presence of a decode \ac{PLA} storing microinstructions. The Intel i960 used microcode to implement several instructions~\cite{intel:i960}.
Another noteworthy \ac{CPU} is the EAL 7 certified AAMP7G by Rockwell Collins~\cite{book:design-verification-microprocessor-systems}. Its separation kernel microcode to realize \ac{MILS} is accompanied with a formal proof.

\subsection{Future Work}
In future work we aim to further explore the microarchitecture and its security implications on system security. We want to highlight microcode capabilities and foster the security and computer architecture communities to incorporate this topic into their future research. We require further knowledge of implemented microarchitectures and update mechanisms to address both attack- and defense-driven research. For example, an open-source \ac{CPU} variant for the security community can lead to instrumentation frameworks and system defenses based on performant microprograms.

\section{Conclusion}


In this paper we successfully changed the behavior of common, general-purpose \acp{CPU} by modification of the microcode. We provided an in-depth analysis of microcode and its update mechanism for AMD K8 and K10 architectures.
In addition, we presented what can be accomplished with this technology:
First, we showed that augmenting existing instructions allows us to implement \ac{CPU}-assisted instrumentation, which can enable high-performance defensive solutions in the future.
Second, we demonstrated that malicious microcode updates can have security implications for software systems running on the hardware.

\section*{Acknowledgement}
We thank the reviewers for their valuable feedback.
Part of this work was supported by the European Research
Council (ERC) under the European Union’s Horizon 2020
research and innovation programme (ERC Starting Grant No.
640110 (BASTION) and ERC Advanced Grant No. 695022 (EPoCH)).
In addition, this work was partly supported by the German Federal Ministry of Education and Research (BMBF Grant 16KIS0592K HWSec).

\section*{Responsible Disclosure} We contacted AMD in a responsible disclosure
process more than 90 days prior to publication and provided detailed information about
our findings.

\bibliographystyle{acm}
{\footnotesize
\bibliography{bibliography}

\begin{thebibliography}{10}

\bibitem{rfc7748}
{\sc {A. Langley \etal}}.
\newblock {Elliptic Curves for Security}.
\newblock {RFC} 7748, {RFC Editor}, January 2016.

\bibitem{amd:revisionguide}
{\sc {Advanced Micro Devices, Inc.}}
\newblock {AMD Athlon\textsuperscript{\textregistered} Processor Model 10
  Revision Guide}, 2003.

\bibitem{amd:pacifica}
{\sc {Advanced Micro Devices, Inc.}}
\newblock {AMD64 Virtualization Codenamed “Pacifica” Technology - Secure
  Virtual Machine Architecture Reference Manual}, 2005.

\bibitem{amd:errata}
{\sc {Advanced Micro Devices, Inc.}}
\newblock {Revision Guide for AMD Family 16h Models 00h-0Fh Processors}, 2013.

\bibitem{book:1976:agrawala}
{\sc Agrawala, A.~K., and Rauscher, T.~G.}
\newblock {\em {Foundations of Microprogramming : Architecture, Software, and
  Applications}}.
\newblock Academic Press, 1976.

\bibitem{link:2004:opteron_exposed}
{\sc Anonymous}.
\newblock {Opteron Exposed: Reverse Engineering AMD K8 Microcode Updates}.
\newblock {[Online]. Available:
  \url{http://www.securiteam.com/securityreviews/5FP0M1PDFO.html}}, 2004.

\bibitem{rompar}
{\sc ApertureLabsLtd}.
\newblock {Semi-automatic extraction of data from microscopic images of Masked
  ROM.}
\newblock \url{https://github.com/ApertureLabsLtd/rompar}.

\bibitem{micro:1999:austin}
{\sc Austin, T.~M.}
\newblock {DIVA: A Reliable Substrate for Deep Submicron Microarchitecture
  Design}.
\newblock In {\em Proceedings of {IEEE/ACM} International Symposium on
  Microarchitecture, {MICRO} 32\/} (1999), pp.~196--207.

\bibitem{ctrsa:2012:brumley}
{\sc {B. B. Brumley \etal}}.
\newblock {Practical Realisation and Elimination of an ECC-Related Software Bug
  Attack}.
\newblock In {\em {CT-RSA}\/} (2012), pp.~171--186.

\bibitem{pkc:2006:bernstein}
{\sc Bernstein, D.~J.}
\newblock {Curve25519: New Diffie-Hellman Speed Records}.
\newblock In {\em PKC\/} (2006), pp.~207--228.

\bibitem{crypto:2008:biham}
{\sc Biham, E., Carmeli, Y., and Shamir, A.}
\newblock {Bug Attacks}.
\newblock In {\em CRYPTO\/} (2008), pp.~221--240.

\bibitem{blazakis2010interpreter}
{\sc Blazakis, D.}
\newblock Interpreter exploitation.
\newblock In {\em USENIX Workshop on Offensive Technologies (WOOT)\/} (2010).

\bibitem{eurocrypt:1997:boneh}
{\sc Boneh, D., DeMillo, R.~A., and Lipton, R.~J.}
\newblock {On the Importance of Checking Cryptographic Protocols for Faults
  (Extended Abstract)}.
\newblock In {\em {EUROCRYPT}\/} (1997), pp.~37--51.

\bibitem{usenix:2003:brumley}
{\sc Brumley, D., and Boneh, D.}
\newblock Remote timing attacks are practical.
\newblock In {\em USENIX Security Symposium\/} (2003).

\bibitem{tr:2014:chen}
{\sc Chen, D.~D., and Ahn, G.-J.}
\newblock {Security Analysis of x86 Processor Microcode}.
\newblock {[Online]. Available:
  \url{https://www.dcddcc.com/docs/2014_paper_microcode.pdf}}, 2014.

\bibitem{isca:2003:corliss}
{\sc Corliss, M.~L., Lewis, E.~C., and Roth, A.}
\newblock {DISE: A Programmable Macro Engine for Customizing Applications}.
\newblock In {\em International Symposium on Computer Architecture\/} (2003),
  pp.~362--373.

\bibitem{iacr:2016:86}
{\sc Costan, V., and Devadas, S.}
\newblock {Intel SGX Explained}.
\newblock Cryptology ePrint Archive, Report 2016/086, 2016.
\newblock {[Online]. Available: \url{http://eprint.iacr.org/2016/086}}.

\bibitem{usenix:2016:costan}
{\sc Costan, V., Lebedev, I., and Devadas, S.}
\newblock {Sanctum: Minimal Hardware Extensions for Strong Software Isolation}.
\newblock In {\em USENIX Security Symposium\/} (2016), pp.~857--874.

\bibitem{book:design-verification-microprocessor-systems}
{\sc {D. S. Hardin}}.
\newblock {\em {Design and Verification of Microprocessor Systems for
  High-Assurance Applications}}.
\newblock Springer, 2010.

\bibitem{link:2007:raadt}
{\sc de~Raadt, T.}
\newblock {Intel Core 2}.
\newblock openbsd-misc mailing list. [Online]. Available:
  \url{http://marc.info/?l-openbsd-isc&m=118296441702631}, 2007.

\bibitem{vries2003_chip_architect}
{\sc de~Vries}.
\newblock {Understanding the detailed Architecture of AMD's 64 bit Core}.
\newblock
  \url{http://www.chip-architect.com/news/2003_09_21_Detailed_Architecture_of_AMDs_64bit_Core.html}.

\bibitem{esorics:2008:duflot}
{\sc Duflot, L.}
\newblock {CPU Bugs, CPU Backdoors and Consequences on Security}.
\newblock In {\em ESORICS\/} (2008), pp.~580--599.

\bibitem{isca:2005:suh}
{\sc {E. G. Suh \etal}}.
\newblock {AEGIS: Architecture for Tamper-Eevident and Tamper-Resistant
  Processing}.
\newblock In {\em {International Conference on Supercomputing}\/} (2003),
  pp.~160--171.

\bibitem{patent:2002:risc86}
{\sc Favor, J.~G.}
\newblock Risc86 instruction set, Jan.~1 2002.
\newblock US Patent 6,336,178.

\bibitem{http://www.fudzilla.com/home/item/32120-amd-denies-existence-of-nsa-backdoor}
{\sc {Fudzilla Staff}}.
\newblock {AMD denies existence of NSA backdoor }.
\newblock {[Online]. Available:
  \url{http://www.fudzilla.com/32120-amd-denies-existence-of-nsa-backdoor}}.

\bibitem{woar:2006:reis}
{\sc {G. A. Reis \etal}}.
\newblock {Configurable Transient Fault Detection via Dynamic Binary
  Translation}.
\newblock In {\em Workshop on Architectural Reliability\/} (2006).

\bibitem{micro:1974:habib}
{\sc Habib, S.}
\newblock {Microprogrammed Enhancements to Higher Level Languages - an
  Overview}.
\newblock In {\em Workshop on Microprogramming\/} (1974), pp.~80--84.

\bibitem{intel:i960}
{\sc {Intel Corporation}}.
\newblock {i960 VH Processor Developer's Manual}, 1998.

\bibitem{intel:errata}
{\sc {Intel Corporation}}.
\newblock {6th Generation Intel\textsuperscript{\textregistered} Processor
  Family Specification Update}, 2016.

\bibitem{intel:pprospec}
{\sc {Intel Corporation}}.
\newblock {Pentium\textsuperscript{\textregistered} Pro Processor Specification
  Update}, 2016.

\bibitem{Jacob:2007:MSC:1543376}
{\sc Jacob, B., Ng, S., and Wang, D.}
\newblock {\em {Memory Systems: Cache, DRAM, Disk}}.
\newblock Morgan Kaufmann Publishers Inc., 2007.

\bibitem{ieee_cs:1975:jones}
{\sc Jones, L.~H.}
\newblock {A Survey of Current Work in Microprogramming}.
\newblock {\em Computer 8}, 8 (Aug. 1975), 33--38.

\bibitem{link:reverse-engineering-arm1}
{\sc {K. Shirriff}}.
\newblock {Reverse engineering the ARM1 processor's microinstructions }.
\newblock [Online]. Available:
  \url{http://www.righto.com/2016/02/reverse-engineering-arm1-processors.html}.

\bibitem{crypto:1996:kocher}
{\sc Kocher, P.~C.}
\newblock {Timing Attacks on Implementations of Diffie-Hellman, RSA, DSS, and
  Other Systems}.
\newblock In {\em {CRYPTO}\/} (1996), pp.~104--113.

\bibitem{libsodium:reference_implementation}
{\sc libsodium}.
\newblock {[Online]. Available:
  \url{https://github.com/jedisct1/libsodium/tree/master/src/libsodium/crypto_core/curve25519/ref10}}.

\bibitem{asplos:2015:hicks}
{\sc {M. Hicks \etal}}.
\newblock {SPECS: A Lightweight Runtime Mechanism for Protecting Software from
  Security-Critical Processor Bugs}.
\newblock In {\em ASPLOS\/} (2015), pp.~517--529.

\bibitem{ccs:2013:maas}
{\sc {M. Maas \etal}}.
\newblock {{PHANTOM:} practical oblivious computation in a secure processor}.
\newblock In {\em CCS\/} (2013), pp.~311--324.

\bibitem{Dachshund2017}
{\sc Maisuradze, G., Backes, M., and Rossow, C.}
\newblock {Dachshund: Digging for and Securing (Non-)Blinded Constants in JIT
  Code}.
\newblock In {\em Symposium on Network and Distributed System Security
  (NDSS)\/} (2017).

\bibitem{patent:2002:patch_device}
{\sc McGrath, K.~J., and Pickett, J.~K.}
\newblock Microcode patch device, Aug.~27 2002.
\newblock US Patent 6,438,664.

\bibitem{dsn:2008:meixner}
{\sc Meixner, A., and Sorin, D.~J.}
\newblock {Detouring: Translating Software to Circumvent Hard Faults in Simple
  Cores}.
\newblock In {\em {IEEE/IFIP} International Conference on Dependable Systems
  and Networks, {DSN}\/} (2008), pp.~80--89.

\bibitem{micro:1988:melvin}
{\sc Melvin, S., and Patt, Y.}
\newblock {SPAM: A Microcode Based Tool for Tracing Operating Sytsem Events}.
\newblock {\em SIGMICRO Newsl. 19}, 1-2 (June 1988), 58--59.

\bibitem{microcode:amd_microprograms}
{\sc Microprograms}.
\newblock {[Online]. Available: \url{https://github.com/RUB-SysSec/Microcode}}.

\bibitem{tc:1980:rauscher}
{\sc Rauscher, T.~G., and Adams, P.~M.}
\newblock {Microprogramming: {A} Tutorial and Survey of Recent Developments}.
\newblock {\em {IEEE} Trans. Computers 29}, 1 (1980), 2--20.

\bibitem{link:evil_maid}
{\sc Rutkowska, J.}
\newblock {Why do I miss Microsoft BitLocker?}
\newblock {[Online]. Available:
  \url{http://theinvisiblethings.blogspot.de/2009/01/why-do-i-miss-microsoft-bitlocker.html}},
  2009.

\bibitem{link:2015:rutowska}
{\sc Rutkowska, J.}
\newblock {Intel x86 considered harmful}.
\newblock {[Online]. Available:
  \url{https://blog.invisiblethings.org/2015/10/27/x86_harmful.html}}, 2015.

\bibitem{jetcs:2016:quadir}
{\sc {S. E. Quadir \etal}}.
\newblock {A Survey on Chip to System Reverse Engineering}.
\newblock {\em J. Emerg. Technol. Comput. Syst. 13}, 1 (Apr. 2016), 6:1--6:34.

\bibitem{ches:2016:ghandali}
{\sc {S. Ghandali \etal}}.
\newblock {A Design Methodology for Stealthy Parametric Trojans and Its
  Application to Bug Attacks}.
\newblock In {\em {CHES}\/} (2016), pp.~625--647.

\bibitem{iccd:2006:narayanasamy}
{\sc {S. Narayanasamy \etal}}.
\newblock {Patching Processor Design Errors}.
\newblock In {\em {International Conference on Computer Design {ICCD}}\/}
  (2006), pp.~491--498.

\bibitem{micro:2007:sarangi}
{\sc {S. R. Sarangi \textit{et al.}}}
\newblock {Patching Processor Design Errors with Programmable Hardware}.
\newblock {\em {IEEE} Micro 27}, 1 (2007), 12--25.

\bibitem{book:practical-introduction-hardware-software-codesign}
{\sc Schaumont, P.~R.}
\newblock {\em {A Practical Introduction to Hardware/Software Codesign}}.
\newblock Springer, 2010.

\bibitem{sintsov2010jit}
{\sc Sintsov, A.}
\newblock Jit-spray attacks \& advanced shellcode.
\newblock {\em HITBSecConf Amsterdam\/} (2010).

\bibitem{Skorobogatov05semi-invasiveattacks}
{\sc Skorobogatov, S.~P.}
\newblock {\em {Semi-Invasive Attacks -- A New Approach to Hardware Security
  Analysis}}.
\newblock PhD thesis, University of Cambridge, 2005.

\bibitem{link:2012:micropr_history}
{\sc Smotherman, M.}
\newblock {A Brief History of Microprogramming}.
\newblock {[Online]. Available:
  \url{http://ed-thelen.org/comp-hist/MicroprogrammingABriefHistoryOf.pdf}},
  2012.

\bibitem{book:computer_architecture:stallings}
{\sc Stallings, W.}
\newblock {\em {Computer Organization and Architecture: Designing for
  Performance (7th Edition)}}.
\newblock Prentice-Hall, Inc., 2005.

\bibitem{sun:opensparc:t2}
{\sc {Sun Microsystems, Inc.}}
\newblock {OpenSPARC Overview}.
\newblock [Online]. Available:
  \url{http://www.oracle.com/technetwork/systems/opensparc/index.html}.

\bibitem{cav:2005:arons}
{\sc {T. Arons \etal}}.
\newblock {Formal Verification of Backward Compatibility of Microcode}.
\newblock In {\em CAV\/} (2005), pp.~185--198.

\bibitem{cans:2016:kaufmann}
{\sc {T. Kaufmann \etal}}.
\newblock {When Constant-Time Source Yields Variable-Time Binary: Exploiting
  Curve25519-donna Built with {MSVC} 2015}.
\newblock In {\em {CANS}\/} (2016), pp.~573--582.

\bibitem{ieee_dt:2010:tehranipoor}
{\sc Tehranipoor, M., and Koushanfar, F.}
\newblock {A Survey of Hardware Trojan Taxonomy and Detection}.
\newblock {\em IEEE Des. Test 27}, 1 (Jan. 2010), 10--25.

\bibitem{Arrigo:2015:Troopers}
{\sc Triulzi, A.}
\newblock Pneumonia, shardan, antibiotics and nasty mov: a dead hand's tale.
\newblock {[Online]. Available:
  \url{https://www.troopers.de/events/troopers15/449\_pneumonia\_shardan\_antibiotics\_and\_nasty\_mov\_a\_dead\_hands\_tale/}},
  2015.

\bibitem{Arrigo:2016:Troopers}
{\sc Triulzi, A.}
\newblock The chimaera processor.
\newblock {[Online]. Available:
  \url{https://www.troopers.de/events/troopers16/655\_the\_chimaera\_processor/}},
  2016.

\bibitem{mucic:1951:wilkes}
{\sc Wilkes, M.~V.}
\newblock {The Best Way to Design an Automatic Calculating Machine}.
\newblock In {\em The Early British Computer Conferences}. MIT Press, 1989,
  pp.~182--184.

\bibitem{eetimes:1997:wolfe}
{\sc Wolfe, A.}
\newblock {For Intel, it’s a case of FPU all over again}.
\newblock EETimes [Online]. Available:
  \url{http://www.fool.com/EETimes/1997/EETimes970516d.htm}, 1997.

\end{thebibliography}
}

\appendix
\
\newpage

\section{Appendix} 

\subsection{Microcode Specification}
\label{ucode:appendix:microcode_specification}

As explained in Section~\ref{ucode:results:heatmaps}, we designed automated test cases to record which locations of the microcode ROM contain triads used to implement a certain x86 instruction. We then cleared the artefacts caused by our test environment and combined the heat maps of all vector path instructions. Table~\ref{fig:heatmap} shows an excerpt of the result.

\begin{table}[!htb]
    \centering
\begin{tabular}{|l|l|}
\hline
\cellcolor{black!15} ROM Address & \cellcolor{black!15} vector instruction\\
\hline
\texttt{0x900 - 0x913} & \texttt{-}\\
\texttt{0x900 - 0x913} & \texttt{-}\\
\texttt{0x914 - 0x917} & \texttt{rep\_cmps\_mem8}\\
\texttt{0x918 - 0x95f} & \texttt{-}\\
\texttt{0x960        } & \texttt{mul\_mem16}\\
\texttt{0x961        } & \texttt{idiv}\\
\texttt{0x962        } & \texttt{mul\_reg16}\\
\texttt{0x963        } & \texttt{-}\\
\texttt{0x964        } & \texttt{imul\_mem16}\\
\texttt{0x965        } & \texttt{bound}\\
\texttt{0x966        } & \texttt{imul\_reg16}\\
\texttt{0x967        } & \texttt{-}\\
\texttt{0x968        } & \texttt{bts\_imm}\\
\texttt{0x969 - 0x971} & \texttt{-}\\
\texttt{0x972 - 0x973} & \texttt{div}\\
\texttt{0x974 - 0x975} & \texttt{-}\\
\texttt{0x976 - 0x977} & \texttt{idiv}\\
\texttt{0x978        } & \texttt{-}\\
\texttt{0x979 - 0x97a} & \texttt{idiv}\\
\texttt{0x97b - 0x9a7} & \texttt{-}\\
\texttt{0x9a8        } & \texttt{btr\_imm}\\
\texttt{0x9a9 - 0x9ad} & \texttt{-}\\
\texttt{0x9ae        } & \texttt{mfence}\\
\texttt{0x9af - 09ff } & \texttt{-}\\
\hline
\end{tabular}
    \caption{Truncated microcode \ac{ROM} heat map.}
    \label{fig:heatmap}
\end{table}

\newpage

In Section~\ref{ucode:results:ucode_instruction_set} we presented the microcode instruction set structure, which is one major result of our reverse engineering effort. We found four operation classes that separate operations of different domains. The operation type determines the exact operation such as \texttt{add} or \texttt{mul}. Our collection of operation types and their encodings are listed in Table~\ref{fig:op_types}.

\begin{table}[!htb]
    \centering
\begin{tabular}{|c|c|c|}
\hline
\cellcolor{black!15} Op Class & \cellcolor{black!15} Mnem & \cellcolor{black!15} Encoding\\
\hline
RegOp & \texttt{add} & \texttt{000000000}\\
RegOp & \texttt{or} & \texttt{000000001}\\
RegOp & \texttt{adc} & \texttt{000000010}\\
RegOp & \texttt{sbb} & \texttt{000000011}\\
RegOp & \texttt{and} & \texttt{000000100}\\
RegOp & \texttt{sub} & \texttt{000000101}\\
RegOp & \texttt{xor} & \texttt{000000110}\\
RegOp & \texttt{cmp} & \texttt{000000111}\\
RegOp & \texttt{test} & \texttt{000001000}\\
\hline
RegOp & \texttt{rll} & \texttt{000010000}\\
RegOp & \texttt{rrl} & \texttt{000010001}\\
RegOp & \texttt{sll} & \texttt{000010100}\\
RegOp & \texttt{srl} & \texttt{000010101}\\
\hline
RegOp & \texttt{mov} & \texttt{001100000}\\
RegOp & \texttt{mul} & \texttt{001110000}\\
RegOp & \texttt{imul} & \texttt{001110001}\\
\hline
RegOp & \texttt{bswap} & \texttt{111000000}\\
RegOp & \texttt{not} & \texttt{111110101}\\
\hline
SpecOp & \texttt{writePC} & \texttt{001000000}\\
SpecOp & \texttt{branchCC} & \texttt{0101CCCCC}\\
\hline
LdOp & \texttt{ld} & \texttt{001111111}\\
StOp & \texttt{st} & \texttt{101010000}\\
\hline
\end{tabular}
    \caption{Collection of microcode operation types.}
    \label{fig:op_types}
\end{table}

\newpage

The microinstruction structure provides two dedicated register fields. One additional register field can be unlocked by enabling  register mode, which replaces the 16-bit immediate field. The register fields can encode a number of registers including x86 general-purpose registers and microcode registers. The microcode registers cannot be accessed by x86 instructions. The contents of the microcode registers are only persistent while one macroinstruction is decoded. Most of the microcode registers serve as general-purpose space for immediate values. However, special microcode registers exist that hold the next decode program counter (pcd) or always read as zero (zerod). We listed the microcode registers with mnemonics and encoding in Table~\ref{fig:reg_encodings}.

\begin{table}[!htb]
    \centering
\begin{tabular}{|c|c|c|c|c|}
\hline
\multicolumn{4}{|c|}{\cellcolor{black!15}Size} & \cellcolor{black!15}Encoding \\
\hline
\cellcolor{black!15} 00 & \cellcolor{black!15} 01 & \cellcolor{black!15} 10 & \cellcolor{black!15} 11 & \cellcolor{black!15}\ \\
\hline

\texttt{al} & \texttt{ax} & \texttt{eax} & \texttt{rax} & \texttt{000000}\\
\texttt{cl} & \texttt{cx} & \texttt{ecx} & \texttt{rcx} & \texttt{000001}\\
\texttt{dl} & \texttt{dx} & \texttt{edx} & \texttt{rdx} & \texttt{000010}\\
\texttt{bl} & \texttt{bx} & \texttt{ebx} & \texttt{rbx} & \texttt{000011}\\
\texttt{ah} & \texttt{sp} & \texttt{esp} & \texttt{rsp} & \texttt{000100}\\
\texttt{ch} & \texttt{bp} & \texttt{ebp} & \texttt{rbp} & \texttt{000101}\\
\texttt{dh} & \texttt{si} & \texttt{esi} & \texttt{rsi} & \texttt{000110}\\
\texttt{bh} & \texttt{di} & \texttt{edi} & \texttt{rdi} & \texttt{000111}\\
\hline
\texttt{t1l} & \texttt{t1w} & \texttt{t1d} & \texttt{t1q} & \texttt{001000}\\
\texttt{t2l} & \texttt{t2w} & \texttt{t2d} & \texttt{t2q} & \texttt{001001}\\
\texttt{t3l} & \texttt{t3w} & \texttt{t3d} & \texttt{t3q} & \texttt{001010}\\
\texttt{t4l} & \texttt{t4w} & \texttt{t4d} & \texttt{t4q} & \texttt{001011}\\
\texttt{t1h} & \texttt{t5w} & \texttt{t5d} & \texttt{t5q} & \texttt{001100}\\
\texttt{t2h} & \texttt{t6w} & \texttt{t6d} & \texttt{t6q} & \texttt{001101}\\
\texttt{t3h} & \texttt{t7w} & \texttt{t7d} & \texttt{t7q} & \texttt{001110}\\
\texttt{t4h} & \texttt{t8w} & \texttt{t8d} & \texttt{t8q} & \texttt{001111}\\
\hline
\texttt{regmb} & \texttt{regmw} & \texttt{regmd} & \texttt{regmq} & \texttt{101000}\\
\texttt{regb} & \texttt{regw} & \texttt{regd} & \texttt{regq} & \texttt{101100}\\
\texttt{pcb} & \texttt{pcw} & \texttt{pcd} & \texttt{pcq} & \texttt{111000}\\
\texttt{zerob} & \texttt{zerow} & \texttt{zerod} & \texttt{zeroq} & \texttt{111111}\\
\hline
\end{tabular}
    \caption{General-purpose and microcode register encodings.}
    \label{fig:reg_encodings}
\end{table}

\newpage

\subsection{Hardware Analysis}

\begin{figure}[ht]
	\centering
	\resizebox{0.995\linewidth}{!}{\includegraphics[scale=1.2,trim={0 0 0.5cm 0},clip]{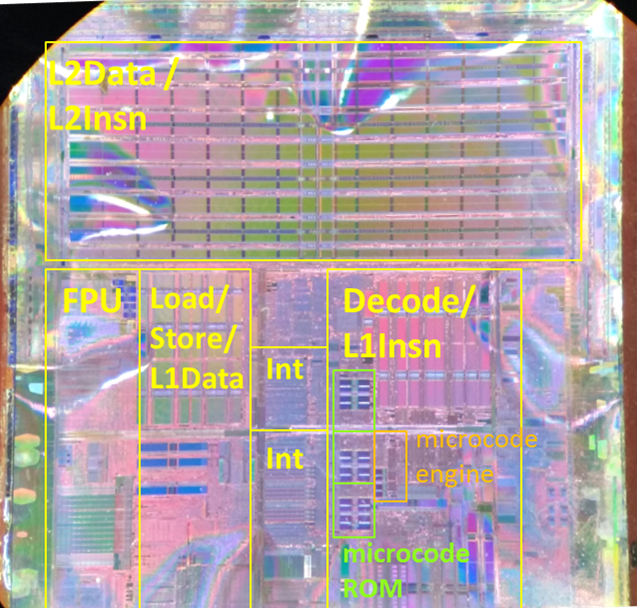}}
	\caption{Die shot of AMD K8 Sempron 3100+ with different \ac{CPU} parts. The image was taken with an optical microscope with low magnification. The die is corrugated due to a remaining thickness below 10 micrometers.}
    \label{fig:cpu_overview}
\end{figure}

In Section~\ref{sec:hw_re} we investigate the hardware of the AMD K8 Sempron 3100+. Hence, we decapsulated and backside-thinned a die to obtain a high-level view of the \ac{CPU} structure. The marked areas are adopted from~\cite{vries2003_chip_architect}, since they show  multiple similarities with our die shot in Figure~\ref{fig:cpu_overview}.
Note that we focus on the microcode \ac{ROM} (marked in green) and neglect the rest of the chip.

\newpage

\subsection{Microprograms}
\label{ucode:appendix:microprograms}

In Section~\ref{ucode:section:microprograms:instrumentation} we present a constructive application of microcode updates, namely program instrumentation. To demonstrate the feasibility, we implemented a proof-of-concept instrumentation that counts the occurrences of the x86 instruction \texttt{div} during execution. It should be noted that the current implementation has some drawbacks, such as reserving two general-purpose registers to steer the instrumentation. However, this is not a fundamental limitation but an engineering issue. The implementation of our proof-of-concept instrumentation is given in Listing~\ref{ucode:listing:instrumentation}.

\begin{minipage}{0.95\linewidth}
\begin{lstlisting}[
    label=ucode:listing:instrumentation,
    aboveskip=1.5em,
    numbers=left, stepnumber=1,numberstyle=\tiny,numbersep=10pt,
    caption={Microprogram that instruments the x86 instruction \texttt{div} and counts the occurrences.}
]
// set match register 0 to 0x7e5

.start 0x0
// load magic constant
mov t1d, 0x0042
sll t1d, 16
add t1d, 0xf00d

// compare and condense
sub t1d, esi
srl t2d, t1d, 16
or t1d, t2d
srl t2d, t1d, 8
or t1d, t2d
srl t2d, t1d, 4
or t1d, t2d
srl t2d, t1d, 2
or t1d, t2d
srl t2d, t1d, 1
or t1d, t2d
and t1d, 0x1

// invert result
xor t1d, 0x1

// conditionally count
ld t2d, [edi]
add t2d, t1d
st [edi], t2d

.sw_branch 0x7e6
\end{lstlisting}
\end{minipage}

\newpage

As explained in Section~\ref{ucode:section:ucode_programs:crypto}, we exploit the x86 \texttt{shrd} instruction to implement both the bug attack and the timing attack. The bug attack in our \ac{RTL} is shown in Listing~\ref{ucode:listing:crypto:ucode_trojan}. Note that in order to hook the \texttt{shrd} instruction, we have to set a match register to the address \texttt{0xaca}. The magic constant as well as the bug value added to the final computation can be arbitrarily chosen.

\begin{minipage}{0.95\linewidth}
\begin{lstlisting}[
    label=ucode:listing:crypto:ucode_trojan,
    aboveskip=1.5em,
    numbers=left, stepnumber=1,numberstyle=\tiny,numbersep=10pt,
    caption={Microprogram that intercepts the x86 instruction \texttt{shrd} and inserts a bug that can be leveraged for a bug attack.}
]
// set match register 0 to 0xaca

.start 0x0
// load magic constant
mov t1d, 0x0042
sll t1d, 16
add t1d, 0xf00d

// compare and condense
sub t1d, esi
srl t2d, t1d, 16
or t1d, t2d
srl t2d, t1d, 8
or t1d, t2d
srl t2d, t1d, 4
or t1d, t2d
srl t2d, t1d, 2
or t1d, t2d
srl t2d, t1d, 1
or t1d, t2d
and t1d, 0x1

// invert result
xor t1d, 0x1

// read immediate
sub t2d, pcd, 0x1
ld t2d, [t2d]
and t2d, 0xff

// implement semantics of shrd
srl regmd4, t2d
mov t3d, 32
sub t3d, t2d
sll t2d, regmd6, t3d
or regmd4, t2d

// conditionally insert bug
add regmd4, t1d

.sw_complete
\end{lstlisting}
\end{minipage}

\newpage

\subsection{Using ASM.JS to remotely trigger a x86 \texttt{div} microcode Trojan}
\label{appendix:subsec:asm.js_trigger}

As explained in Section~\ref{sec:subsec:ucode_programs:remote}, we use ASM.JS
code in Firefox 50 to trigger the implemented x86 \texttt{div} Trojan. It is shown in
Listing~\ref{appendix:listing:asm.js_trigger}. Instead of using \texttt{nop} and
\texttt{int3} instructions, arbitrary payloads can be implemented. For example,
the attacker might deploy a remote shell as soon as the microcode Trojan is
triggered, which establishes a connection to her remote control server.

\begin{minipage}{0.95\linewidth}
\begin{lstlisting}[
    style=JS,
    caption={ASM.JS code within a remote web page which emits a \texttt{div ebx} instruction and an attacker-controlled payload in Firefox 50.0.},
    label=appendix:listing:asm.js_trigger,
]
<!DOCTYPE HTML>
<html>
<script>
/*
Firefox 50.0 32-bit on Linux
We use a non-weaponized payload. Instructions

offset:         opcodes       assembly
=======         =======       ========
0x00000000:     05909090a8    add eax, 0xa8909090
0x00000005:     05909090cc    add eax, 0xcc909090

become a nop-sled with a breakpoint at the
end, if the first instruction is executed
from offset 1:

offset:          opcodes        assembly
=======          =======        ========
0x00000001:      90             nop
0x00000002:      90             nop
0x00000003:      90             nop
0x00000004:      a805           test al, 5
0x00000006:      90             nop
0x00000007:      90             nop
0x00000008:      90             nop
0x00000009:      cc             int3
*/
function generate_microcode_trigger(){
    "use asm";
    function exec_payload(dividend, divisor){
        dividend = dividend|0;
        divisor = divisor|0;
        var val = 0;
        /* div ebx */
        val = ((dividend>>>0)/(divisor>>>0))>>>0;
        /* add eax, 0xA8909090 */
        val = (val + 0xa8909090)|0;
        /* add eax, 0xCC909090 */
        val = (val + 0xcc909090)|0;
        return val|0;
    }
    return exec_payload;
}

function main(){
    /* trigger condition: */
    /* dividend */
    eax = 0xa1a2a3a4
    /* divisor */
    ebx = 0xb1b2b3b4

    trigger_microcode_trojan = generate_microcode_trigger();
    trigger_microcode_trojan(eax, ebx);
}
</script>
<body onload=main()>
</body>
</html>
\end{lstlisting}
\end{minipage}

\end{document}